\documentclass[fleqn,usenatbib]{mnras}

\usepackage{newtxtext,newtxmath}
\usepackage{mathptmx}

\usepackage[T1]{fontenc}
\usepackage[LGRgreek]{mathastext}

\DeclareRobustCommand{\VAN}[3]{#2}
\let\VANthebibliography\thebibliography
\def\thebibliography{\DeclareRobustCommand{\VAN}[3]{##3}\VANthebibliography}

\usepackage{multirow}
\usepackage{subcaption}
\usepackage{orcidlink}
\usepackage{graphicx}	
\usepackage{amsmath}	

\title[Non-Linear Solar EUV-Driven Sodium Release]{Non-Linear Solar EUV-Driven Sodium Release from the Lunar Surface: A Contrast to the Linear PSD Model}

\author[A. Devaraj et al.]{A. Devaraj,$^{1,2}$\thanks{ashish.devaraj@res.christuniversity.in}\orcidlink{0000-0001-5933-058X}
S. Narendranath,$^{3}$\thanks{kcshyama@ursc.gov.in}\orcidlink{0000-0001-9199-4925}
Sreeja. S. Kartha$^{1}$\thanks{sreeja.kartha@christuniversity.in}\orcidlink{0000-0002-7666-1062}
M. Sarantos$^{4}$\orcidlink{0000-0003-0728-2971}
Akhil Krishna R.$^{1}$\orcidlink{0000-0002-6096-3330}\newauthor
Blesson Mathew$^{1}$\orcidlink{0000-0002-7254-191X}
T. Sivarani$^{5}$\orcidlink{0000-0003-0891-8994}
S. Nidhi$^{1,6}$\orcidlink{0000-0003-2825-147X}
P. Anbazhagan$^{5}$
and G. Selvakumar$^{5}$
\\\\
$^{1}$Department of Physics and Electronics, CHRIST (Deemed to be University), Bengaluru, India\\
$^{2}$Aryabhatta Research Institute of Observational Sciences, Devasthal Observatory, Nainital, Uttarakhand, India\\
$^{3}$Space Astronomy Group, U R Rao Satellite Centre, ISRO, Bengaluru, India\\
$^{4}$Heliophysics Science Division, NASA Goddard Space Flight Center Greenbelt, MD, USA\\
$^{5}$Indian Institute of Astrophysics, Bengaluru, India\\
$^{6}$The Oxford College of Science, 17th, 32, 19th Main road, Sector 4, HSR Layout, Bengaluru, Karnataka, India\\
}

\date{Accepted 2025 August 27. Received 2025 August 14; in original form 2024 November 13}
\pubyear{\the\year{}}

\begin{document}
\label{firstpage}
\pagerange{\pageref{firstpage}--\pageref{lastpage}}
\maketitle

\begin{abstract}
The correlation between solar Extreme Ultra-Violet (EUV) radiation above 8.8\,eV and the release of sodium from the lunar surface via photon-stimulated desorption (PSD) is investigated. We use simultaneous measurements of EUV photon flux and Na optical spectral line flux ($F_{Na}$) from the lunar exosphere. Data were acquired with the high-resolution (R$\sim$72000) Echelle Spectrograph on the 2.34-m Vainu Bappu Telescope during the lunar first quarter (January-March 2024), observing Na\,I D2 and D1 flux at altitudes below $\sim$590\,km from the surface. Simultaneous EUV and FUV measurements were acquired from the GOES-R Series Extreme Ultraviolet Sensor (EUVS), while NUV data were obtained from the Total and Spectral Solar Irradiance Sensor-1 (TSIS-1) aboard the ISS. We correlated $F_{Na}$ with EUV photon flux from EUVS across six bands spanning 256-1405\,\AA\ (48.5-8.8\,eV) and NUV (2000-4000\,\AA) from TSIS-1. A non-linear rise in lunar exospheric sodium with increasing EUV and FUV fluxes was observed, contrasting with previous linear PSD models. The EUV radiation above 10\,eV drives sodium release, with 256-304\,\AA\ wavelengths as dominant contributors. Additionally, the NUV flux and $F_{Na}$ are positively correlated, indicating the role of sodium release. The zenith column density averages $3.3\times10^{9}$ atoms cm$^{-2}$, with Characteristic temperatures averaging at $\sim$6700\,K and scale heights of $\sim$1500\,km. Elevated temperatures and sodium densities during solar activity suggest enhanced Na release during flares. These results emphasize the need for a revised PSD model above 8.8\,eV and improved constraints on the PSD cross-section.
\end{abstract}



\section{Introduction}
Our Moon hosts a tenuous Surface Boundary Exosphere (SBE), which is highly dynamic, quasi-collisionless, and intrinsically non-thermal \citep{Leblanc2022, Milillo2023}. These characteristics allow various compositional components to be treated as independent exospheres \citep{Stern1999, Stern2012}. Several studies have reported the presence of noble gases \citep{Hoffman1973, Hodges1974, benson1975, Nakamura1977, Das2016, Thampi2015, Stern2012, dhanya2021}, volatile elements \citep{potter1988a, Tyler1988}, and molecules such as ammonia (NH${3}$) \citep{hoffman1974}, methane (CH${4}$) \citep{hoffman1974,Hodges2016}, and carbon dioxide (CO$_{2}$) \citep{hoffman1975,Stern1999} in the lunar SBE. Moderately volatile elements such as sodium (Na) and potassium (K) were first observed in the SBE through ground-based spectroscopy \citep{potter1988a, Potter1988b, Tyler1988} and imaging \citep{mendillo1993}. Although Na is a minor element in the lunar regolith \citep{Narendranath2022}, most studies have focused on the Na\,I - D1 (5895.93\,\text{\AA}) and D2 (5889.96\,\text{\AA}) lines because Na is easily observable with ground-based telescopes due to its high resonance scattering properties.

The dynamic Na exosphere is sustained through a complex interplay of various processes, including sources, sinks, and escape mechanisms \citep{Teolis2023}. Source processes generating the exosphere include (i) photon-stimulated desorption (PSD) by solar UV radiation \citep{Yakshinskiy1999, sarantos2010}, (ii) solar wind ion sputtering \citep{smith2001, crider2003, wurz2007, milillo2011}, (iii) micrometeorite impact vapourization (MIV) \citep{verani1998, szalay2016, janches2021, berezhnoy2014, killen2019}, (iv) radioactive decay \citep{hodges1977b, Kockarts1973, killen2002}, and (v) thermal desorption \citep{Dukes2017}. Gravitational effects govern the sink processes, while escape mechanisms include gravitational escape, solar radiation pressure, and loss to the solar wind via photoionization \citep{Killen2018}. Together, these mechanisms maintain the tenuous yet dynamic exosphere surrounding the Moon, which can be studied through optical spectroscopic observations.

Previous observations reveal a two-component structure within the lunar exosphere, driven by various release processes \citep{Kuruppuaratchi2018}. The first component consists of a hot population of recently released Na atoms with higher characteristic temperatures ($T_{ch}$), produced by PSD, MIV, and charged particle sputtering \citep{Ip1991, morgan_and_shemansky1991}. The second component is a cooler population formed through the collisional thermalization of the hot atoms as they interact with the lunar surface \citep{Potter1988b, potter_and_morgan1991, sprague1992}. By measuring the Doppler broadening of Na lines, we can infer the $T_{ch}$ of these atoms in the exosphere. However, in a collision-less atmosphere, the atoms deviate from a maxwellian velocity distribution since the atoms are not collisionally thermalized in flight. Hence, the temperatures obtained from doppler measurements are only considered to be the representative temperature of the atoms in the exosphere \citep{Kuruppuaratchi2018, kuruppuaratchi2023, Mierkiewicz2014}. In addition, the assumption that the lunar exosphere is a collision-less atmosphere is valid for processes like PSD, solar wind sputtering, and thermal desorption, as these act on individual atoms. However, MIV may deviate from this assumption, as collisional thermalization in the impact-generated plume is possible, allowing particles to achieve a maxwellian velocity distribution \citep{Eichhorn1978, mangano2007} PSD and solar wind ion sputtering are the dominant processes, while micrometeorites become significant episodically \citep{sarantos2010, verani1998, szalay2016, janches2021, berezhnoi2023}. Thermal desorption signatures are evident near the surface with scale heights less than 80\,km and temperatures below $\sim$400\,K \citep{sprague1992}. Release due to radioactive decay is localized based on surface abundance \citep{hodges1980, killen2002}.

Among all the processes, PSD is identified as the predominant source process in subsolar regions, while other processes dominate near the terminator \citep{mendillo1999, Kuruppuaratchi2018, kuruppuaratchi2023}. PSD is a non-thermal process induced by UV photons from the Sun, which causes electronic transitions in surface minerals. The mineral surface holds Na in its ionic state as an adsorbate layer. The electronic transitions in the bulk material result in free electrons entering the conduction band. In its adsorbed state, Na captures these free conduction electrons, increasing in size from $1.4\,\text{\AA}$ to $2.8\,\text{\AA}$, becoming a neutral Na atom \citep{Dukes2017}. This size change generates an anti-bonding repulsive state, releasing Na into the exosphere \citep{Mandey1998, Yakshinskiy1999}. Photon energies that cause PSD are in the range of 4-10 eV \citep{Wurz2022, Yakshinskiy1999}, with \cite{Yakshinskiy1999} noting that the PSD cross-section increases for photon energies exceeding 5 eV and estimate a Na source rate of $4\times10^{6}$\,atoms $cm^{-2}$ $s^{-1}$. They also quote that the solar photon flux reaching the surface decreases quickly by two orders of magnitude or higher at photon energies $>$8\,eV.

Our study uniquely investigates the effect of Extreme Ultra-Violet (EUV) radiation from the Sun, specifically in the photon energy range of 8.8\,eV to 48.5\,eV, on the release of Na from the lunar surface. This energy range extends beyond the previously recognized photon energy range associated with PSD. We compared simultaneous EUV and Far Ultra-Violet (FUV) flux measurements from the Geostationary Operational Environmental Satellite (GOES)-R Extreme Ultraviolet Sensor (EUVS) with optical high-resolution spectroscopic line strengths of the Na\,I D1 and D2 lines obtained using the Vainu Bappu Telescope (VBT). Additionally, the variation of Na with Near Ultra-Violet (NUV) irradiance from the Sun was observed using data (2000-4000\,\text{\AA}) from the Total and Spectral Solar Irradiance Sensor-1 (TSIS-1) instrument aboard the International Space Station (ISS). We also measured the $T_{ch}$ and scale heights of the exosphere within $\sim$590\,km of the lunar surface to identify the presence of a non-thermal component in the lunar exosphere. Since the observations carried out in our study are only near the first quarter Moon phase with $\sim$12\% variation in illumination, it offers the advantage of minimizing the effects of monthly variation. However, seasonal variation mentioned in \cite{Colaprete2016} may still be expected.

The paper is organized as follows: Section \ref{sec:observation_and_data} describes the Lunar exospheric observations carried out and the data used in this work. Section \ref{sec:reduction_and_analysis} details the analysis procedure, including how the exospheric spectra were extracted from the observed spectra. The results and discussion are presented in Sections \ref{sec:results} and \ref{sec:descussion}, respectively, followed by the conclusion in Section \ref{sec:conclusion}.

\begin{figure}
    \centering
    \includegraphics[width=\columnwidth]{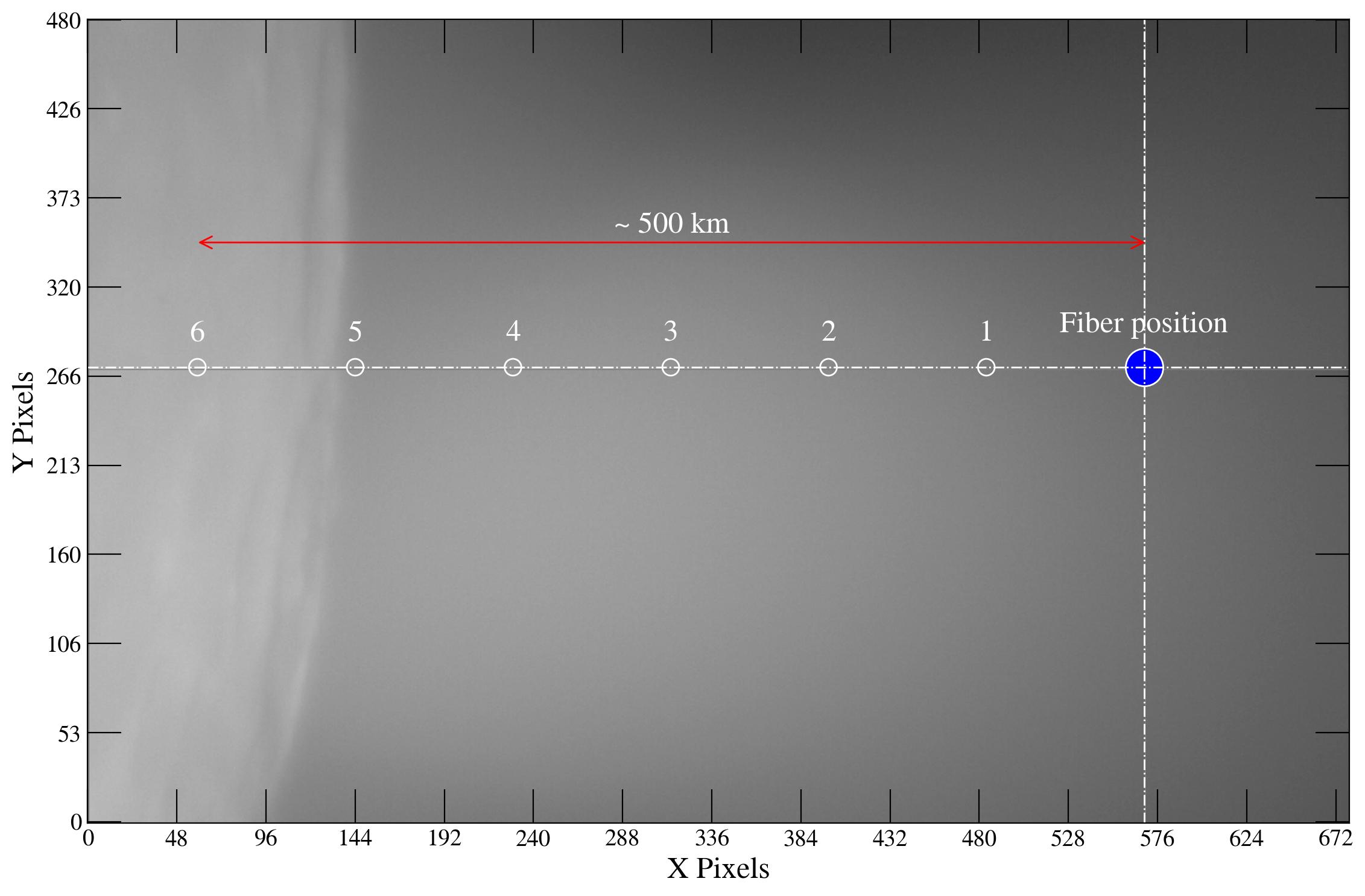}
    \caption{The output image of the Eastern limb of the Moon from the IDS camera attached to the guide telescope of VBT is shown in the Figure. The X-axis spans 720 pixels ($\sim5'$), and the Y-axis spans 480 pixels ($\sim3'$). The representative blue-filled circle at $\sim$(570, 270) pixels shows the fibre (diameter of 2.7$''$) position of the main telescope. The apparent edge of the surface of the Moon was shifted from positions 1 to 6, and limb spectra were acquired for each height. The height measured between position 6 and the fibre is $\sim$500\,km.}
    \label{fig:IDS_camera_output}
\end{figure}

\begin{table*}
\caption{Log of observations carried out at VBT. The astrometric right ascension and declination of the Moon's centre with respect to the observing site in the reference frame of the planetary ephemeris (ICRF) are given, along with the local hour angle. The illumination percentage, phase angle and exposure time are also provided for each observation.}             
\label{table:observations}      
\centering          
\begin{tabular}{lccccccc}     
\hline\hline       
\noalign{\smallskip}
Date& MJD &  \multirow{2}{*}{\begin{tabular}[c]{@{}c@{}}Illumination\\ (\%)\end{tabular}}  &  \multirow{2}{*}{\begin{tabular}[c]{@{}c@{}}Phase angle\\ (deg)\end{tabular}}  &  \multirow{2}{*}{\begin{tabular}[c]{@{}c@{}}Exp Time\\ (s)\end{tabular}} & \multirow{2}{*}{\begin{tabular}[c]{@{}c@{}}RA\\ (hh:mm:ss)\end{tabular}} & \multirow{2}{*}{\begin{tabular}[c]{@{}c@{}}DEC\\ (dd:mm:ss)\end{tabular}} & \multirow{2}{*}{\begin{tabular}[c]{@{}c@{}}Local HA\\ (hh:mm:ss)\end{tabular}} \\ \\ \hline
\noalign{\smallskip}
17 January 2024 & 60326.607 & 43.46 & 82.9 & 300 & 01:09:48.61 & +07:33:40.5 & +02:26:09 \\
& 60326.621 & 43.56 & 83.07 & 900 & 01:10:13.53 & +07:38:24.6 & +02:44:40 \\
& 60326.649 & 43.77 & 83.44 & 1200 & 01:11:09.44 & +07:48:12.5 & +03:24:44 \\
\noalign{\smallskip}
\noalign{\smallskip}
18 January 2024 & 60327.568 & 54.48 & 95.15 & 900 & 01:59:50.42 & +13:33:09.2 & +00:43:26 \\
& 60327.657 & 55.05 & 96.26 & 1200 & 02:02:27.31 & +14:00:45.9 & +02:48:25 \\
& 60327.672 & 55.17 & 96.46 & 1200 & 02:02:59.84 & +14:05:20.4 & +03:10:41 \\
\noalign{\smallskip}
\noalign{\smallskip}
16 February 2024 & 60356.578 & 49.38 & 89.42 & 600 & 03:31:54.37 & +22:43:00.6 & +01:20:08 \\
& 60356.588 & 49.44 & 89.53 & 1200 & 03:32:11.62 & +22:44:47.7 & +01:33:22 \\
& 60356.614 & 49.6 & 89.85 & 1800 & 03:33:03.93 & +22:49:37.0 & +02:10:10 \\
& 60356.661 & 49.92 & 90.43 & 1800 & 03:34:45.25 & +22:56:48.9 & +03:16:10 \\ 
\noalign{\smallskip}
\noalign{\smallskip}
16 March 2024 & 60385.586 & 43.67 & 82.97 & 1200 & 05:08:54.41 & +27:53:48.2 & +01:49:27 \\
\noalign{\smallskip}
\noalign{\smallskip}
17 March 2024 & 60386.661 & 54.61 & 95.72 & 1200 & 06:10:44.87 & +28:42:11.2 & +02:39:20 \\
& 60386.693 & 54.82 & 96.09 & 1200 & 06:12:01.02 & +28:38:24.0 & +03:23:07 \\

\hline                  
\end{tabular}
\end{table*}
\section{Observation and Data}\label{sec:observation_and_data}
In this work, we utilize high-resolution (R$\sim$72,000) optical spectroscopic data in the wavelength range $\sim$3000 - 9000\,\text{\AA} from VBT, as well as EUV and FUV data ($\sim$256 - 1405\,\text{\AA}) from GOES-R \citep{eparvier2009}. Hereafter, these two bands are collectively referred to as $F_{GOES}$. Additionally, we incorporate NUV data (2000 - 4000\,\text{\AA}) from the TSIS-1 instrument \citep{richard2020}.

\subsection{Optical spectroscopic data}
The data for this work were acquired using the High-Resolution Fibre-fed Echelle Spectrograph on the 2.34-metre VBT \citep{bhattacharyya1992}. The spectrograph is equipped with a UKATC CCD with a resolution of 4096 x 4096 pixels, each measuring 12 x 12 microns. The CCD operates with a gain of 0.85$e^{-}$ per ADU and a 4.4$e^{-}$ read noise level. The optical fibre core has an aperture of 100\,$\mu$m in diameter, corresponding to an angular size of approximately $2.7''$. The input end of the fibre is located at the Prime Focus of the telescope, where the plate scale is $27''$ per mm. The instrument achieves a spectral resolution of R $\sim$72000, using a 60-micron slit width. Details regarding the optical design and performance of the spectrograph can be found in \cite{Krao2005}.

Observations were conducted on five nights near the first quarter of the Moon in January, February and March 2024, as listed in Table \ref{table:observations}. The excessive brightness of the Moon $(\sim$-10\,mag) was limited by employing a 60 $\mu$m slit to limit the light entering the spectrograph and also to obtain R$\sim$72000. In addition, exposure times only within 300 to 1800\,seconds were used to avoid saturation of the CCD. This setup allowed the acquisition of spectra from both the limb and the disk of the Moon. The limb was sampled at various tangential heights above the apparent surface edge, with observations covering a range within approximately 590 km, accounting for a maximum drift of $\sim$90\,km during 1800\,second exposures. The excessive brightness also prevented using the auto-guider CCD on the VBT for tracking during observations. Hence, an additional guiding camera was installed for manual tracking. For this purpose, an Imaging Development Systems (IDS) camera with a 720 x 480-pixel detector was attached to the 8-inch Guider telescope. This camera operates at a frame rate of 59 fps and covers a temperature range from -20 to +60$^{\circ}$C. The field of view provided by the guider was estimated to be $5.3\pm0.03'$ in the X direction and $3.4\pm0.02'$ in the Y direction. Figure \ref{fig:IDS_camera_output} shows the Eastern equatorial lunar limb region where the spectroscopic data were obtained. Additionally, spectra from three different locations on the disk were observed to create a median disk spectrum. A solar spectrum was obtained during the daytime of observation for aperture extraction while reducing Echelle data. The Th-Ar lamp spectrum was used to calibrate the wavelength of the data. HR3982 was observed as the telluric standard star to remove the Earth's atmospheric component.

As mentioned, the exosphere was sampled at various altitudes up to $\sim$590\,km, a limit set by the field of view (FOV) of the guider CCD used for manual guiding. Altitude estimation was constrained by telescope flexure and encoder inaccuracies. Observing beyond 590 km proved challenging, as manual guiding relied on selecting a clearly visible lunar crater, which was difficult when craters were not within the guider FOV.
To overcome this, we adopted the following observation strategy: The optical fibre of the telescope was positioned at the blue-marked point in the guider CCD view (Figure 1). The guider FOV was then divided into six sections, and spectra were recorded by aligning the apparent lunar edge with each position from 1 to 6. Figure 1 illustrates the configuration when the Moon was placed at position 5 to obtain spectrum. The total observed distance of $\sim$500\,km was estimated between the blue point and position 6 using the pixel scale of the CCD. Additionally, a drift in the fibre position from the marked blue point was observed due to telescope flexure, which varied with the hour angle and declination of the Moon. Encoder-induced drifts further affected positioning, making accurate drift estimation difficult. Longer exposure times increased these drifts, with a maximum observed shift of $\sim$45'' (corresponding to $\sim$90\,km in altitude) for 1800\,second exposures. Given these limitations, we do not provide precise altitude estimates but instead report a maximum observation altitude of $\sim$590 km.

\begin{figure*}
\centering
\includegraphics[width=2\columnwidth]{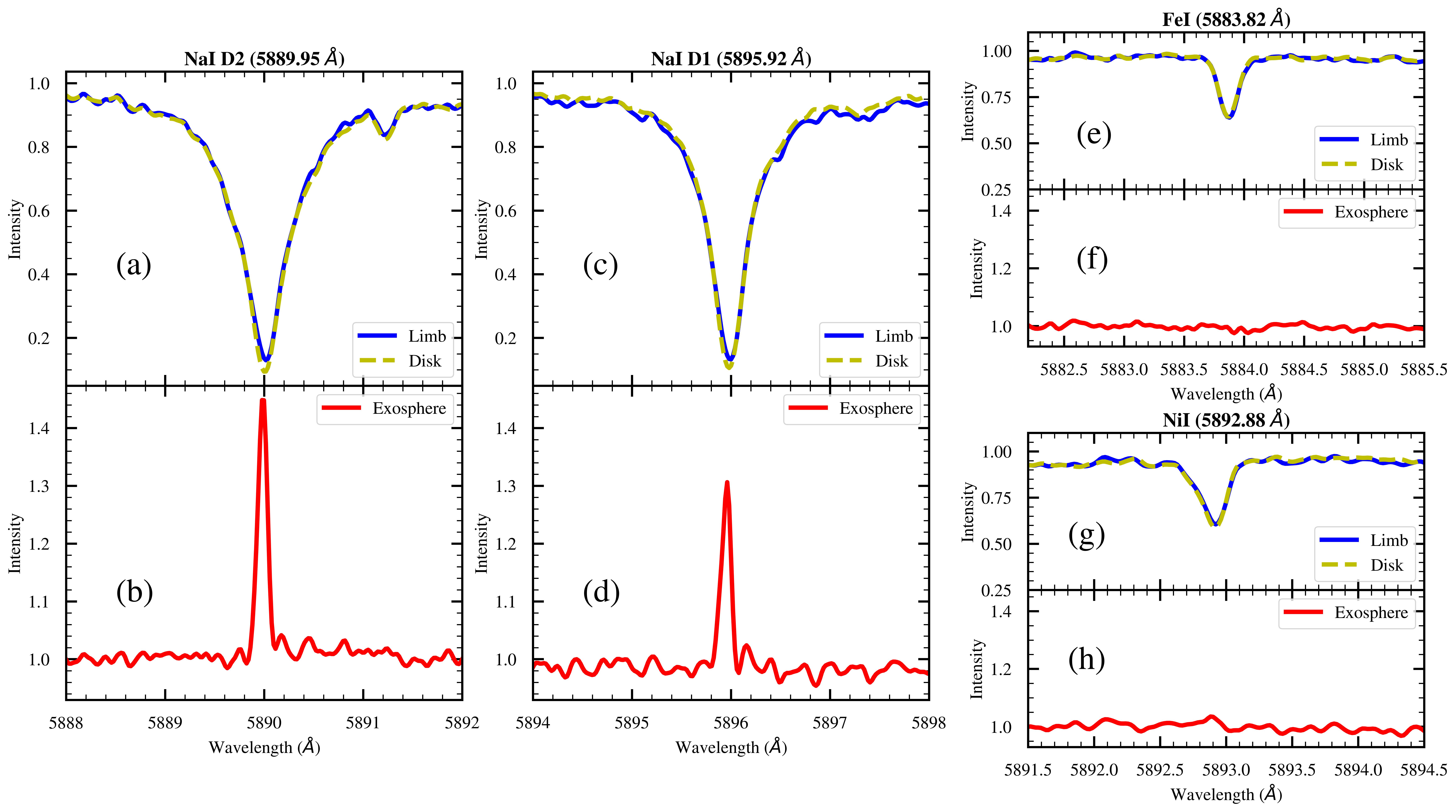}
\caption{The figure displays spectra observed across different regions of the lunar environment during observations on March 16, 2024. The panels (a), (c), (e) and (g) depict the continuum-normalized spectra of the Lunar limb above the surface at the apparent subsolar point (blue solid line) and the median disk spectra obtained from lunar disk observations (yellow dashed line). These panels highlight prominent absorption features such as Na\,I D2, Na\,I D1, Fe\,I, and Ni\,I lines. The panels (b), (d), (f) and (h) show the spectrum of the Lunar exosphere, derived by dividing the median disk spectra from the limb spectra. Among these, panels (b) and (d) display the emission peaks of Na\,I D2 and D1 lines from the Lunar exosphere. The Fe\,I and Ni\,I lines expected only from solar scatter are absent in the exosphere spectrum shown in panels (f) and (h) due to their cancellation during the division process, indicating the removal of the solar scatter component.}
\label{fig:NaExosphere}%
\end{figure*}

\subsection{Ultra-Violet data}
The GOES-R Series EUVS is a key operational space weather instrument that continuously measures solar irradiance at short wavelengths where the output of the Sun varies the most \citep{ermolli2013, goesseries2020}. The EUVS, a part of Extreme Ultraviolet and X-ray Irradiance Sensors \citep{eparvier2009, snow2009}, conducts high spectral-resolution measurements in the EUV and FUV ranges,  capturing distinct solar emission lines that represent different layers of the solar atmosphere. The EUVS can observe various spectral lines across different regions of the Sun. It measures HeII at 256.32\,\text{\AA} and 303.78\,\text{\AA} from the transition region, FeXV at 284.15\,\text{\AA} from the corona and CIII at 1175\,\text{\AA} from the chromosphere. Additionally, CII at 1335.7\,\text{\AA} from the chromosphere, and a line at 1405\,\text{\AA} from the transition region showing SiIV and OIV. We use the science-quality, one-minute averaged Level 2 products available in netCDF format from the GOES-R archive\footnote{\url{https://www.ngdc.noaa.gov/stp/satellite/goes-r.html}} to obtain simultaneous $F_{GOES}$ measurements for our VBT echelle spectroscopic observations. All the $F_{GOES}$ measurements are corrected for the variation in Sun-Moon distance during our observations.\\
\\
Data from the Spectral Irradiance Monitor (SIM) instrument on the Total and Spectral Solar Irradiance Sensor-1 (TSIS-1) aboard the International Space Station (ISS) was used to obtain the integrated NUV (2000-4000\,\text{\AA}) irradiance \citep{richard2019}. SSI provides the solar irradiance spectra with spectral coverage from 2000\,\text{\AA} to 24,000\,\text{\AA} at variable ($\sim$1 to 35\,nm) spectral resolution \citep{richard2020}. The flux from the daily average Level-3 product between the NUV range was integrated and used in the analysis. There are temporal gaps observed in the TSIS-1 SIM data record due to ISS operational activities. Due to this, we obtain simultaneous data on each observation date except for 16 February 2024, when the data is not available. All NUV irradiance measurements are corrected for the variation in Sun-Moon distance during our observations. TSIS-1 is developed by the Laboratory for Atmospheric and Space Physics (LASP). These data are available from the TSIS-1 website\footnote{\url{https://lasp.colorado.edu/tsis/data/}}. These data were accessed via the LASP Interactive Solar Irradiance Datacenter (LISIRD\footnote{\url{https://lasp.colorado.edu/lisird/}}).

\section{Optical data reduction and Analysis} \label{sec:reduction_and_analysis}

\begin{table*}
\caption{This table provides the MJD of each observation, along with the fluxes of the Na\,I D2 and Na\,I D1 spectral lines, reported in kR, and their corresponding flux ratios. Additionally, it includes simultaneous EUVS flux measurements from GOES in EUV and FUV wavelengths for each MJD, along with the daily average NUV flux from TSIS-1. The NUV flux from TSIS-1 was not available from 60356.57816 to 60356.661}
\label{table:flux_measurements}
\centering  
\begin{tabular}{lccccccccccc}
\hline\hline
\noalign{\smallskip}
\multirow{2}{*}{MJD}& \multirow{2}{*}{\begin{tabular}[c]{@{}c@{}}Na\,I D2\\ (kR)\end{tabular}} & \multirow{2}{*}{\begin{tabular}[c]{@{}c@{}}Na\,I D1\\ (kR)\end{tabular}} & \multirow{2}{*}{\begin{tabular}[c]{@{}c@{}}Ratio\\ (D2/D1)\end{tabular}} 
 & \multicolumn{6}{c}{ Photon flux ($\times10^{9}$ $cm^{-2} s^{-1}$)} & \multirow{2}{*}{\begin{tabular}[c]{@{}c@{}}NUV flux\\(W $m^{-2}$ $nm^{-1}$)\end{tabular}} \\ \cline{5-10} 
\noalign{\smallskip}
& & & & 256 \text{\AA}& 284 \text{\AA}& 304 \text{\AA}& 1175 \text{\AA}& 1335 \text{\AA}& 1405 \text{\AA}& \\ \hline
\noalign{\smallskip}
60326.607 & 6.13$\pm$1.23 & 3.31$\pm$0.67 & 1.85$\pm$0.53 & 2.264 & 2.087 & 10.108 & 6.828 & 19.419 & 12.98 & 88442.317 \\
60326.621 & 5.89$\pm$1.18 & 2.97$\pm$0.6 & 1.99$\pm$0.56 & 2.258 & 2.082 & 10.091 & 6.835 & 19.361 & 12.931 & \\
60326.649 & 5.51$\pm$1.11 & 2.97$\pm$0.6 & 1.86$\pm$0.53 & 2.246 & 2.07 & 10.048 & 6.803 & 19.331 & 12.862 & \\
\noalign{\smallskip}
\noalign{\smallskip}
60327.568 & 4.49$\pm$0.9 & 2.87$\pm$0.58 & 1.56$\pm$0.45 & 2.167 & 2.018 & 10.056 & 6.793 & 19.277 & 12.702 & 88450.977 \\
60327.657 & 4.14$\pm$0.83 & 3.24$\pm$0.65 & 1.28$\pm$0.36 & 2.166 & 2.029 & 10.0 & 6.739 & 19.226 & 12.667 & \\
60327.672 & 3.97$\pm$0.8 & 3.51$\pm$0.71 & 1.13$\pm$0.32 & 2.174 & 2.029 & 10.039 & 6.72 & 19.307 & 12.723 & \\
\noalign{\smallskip}
\noalign{\smallskip}
60356.578 & 3.27$\pm$0.66 & 1.87$\pm$0.38 & 1.75$\pm$0.5 & 2.107 & 1.985 & 9.58 & 6.521 & 18.692 & 12.5 & - \\
60356.588 & 3.73$\pm$0.75 & 2.61$\pm$0.53 & 1.43$\pm$0.41 & 2.107 & 1.982 & 9.565 & 6.543 & 18.597 & 12.437 & \\
60356.614 & 4.26$\pm$0.86 & 2.08$\pm$0.42 & 2.05$\pm$0.59 & 2.109 & 1.966 & 9.548 & 6.509 & 18.604 & 12.377 & \\
60356.661 & 2.9$\pm$0.58 & 1.87$\pm$0.38 & 1.55$\pm$0.44 & 2.114 & 1.936 & 9.536 & 6.603 & 18.666 & 12.405 & \\
\noalign{\smallskip}
\noalign{\smallskip}
60385.586 & 3.26$\pm$0.65 & 1.34$\pm$0.27 & 2.44$\pm$0.7 & 1.851 & 1.626 & 8.503 & 5.898 & 16.765 & 11.105 & 88316.363 \\
\noalign{\smallskip}
\noalign{\smallskip}
60386.661 & 3.45$\pm$0.69 & 1.41$\pm$0.28 & 2.46$\pm$0.7 & 1.898 & 1.68 & 8.644 & 6.07 & 17.063 & 11.758 & 88300.363\\
60386.693 & 2.81$\pm$0.57 & 1.42$\pm$0.29 & 1.98$\pm$0.57 & 1.893 & 1.678 & 8.618 & 6.069 & 17.081 & 11.694 & \\
\hline
\end{tabular}
\end{table*}

The data reduction of the obtained spectra was carried out using IMRED, CCDPROC, and ECHELLE packages in IRAF \citep{Tody1986}. The standard data reduction operations of bias subtraction, order identification, scattering correction, flat fielding and spectrum extraction \citep{willmarth1994, churchill1995, aoki2008} were carried out. The wavelength calibration of the spectra was done using a Th-Ar lamp spectrum obtained during each observation night. The final reduced spectra had a spectral dispersion of $\sim0.02''$ per pixel. Further analysis for extracting the exospheric spectra was subsequently carried out using these wavelength-calibrated spectra.\\
\\
\subsection{Exospheric spectra}
The extraction of exospheric spectra involved normalizing the continuum, telluric contamination, and solar scatter from the lunar surface. Initially, spectra were continuum-normalized using Python code and IRAF routines. A univariate spline of third order with a smoothing factor 1.0, applied through the SciPy package \citep{Scipy2020}, was used for continuum fitting. Telluric and interstellar contamination was removed using the B-star correction technique. Additionally, spectra from approximately $\sim40''$ inward of the lunar disk were observed. These spectra represent the solar scatter component from the surface and were used to correct for solar scatter. Figure \ref{fig:NaExosphere} shows the reduced spectra from different regions.

\subsubsection{Telluric and Interstellar contamination}
The spectra of the Moon's limb and disk display absorption features from Earth's atmosphere and the interstellar medium beyond the Moon. These features vary depending on the observation time, location, and atmospheric conditions, potentially contaminating the optical spectra near $\sim$5900\,\text{\AA} with spectral features other than Lunar origin.
HR3982, a rapidly rotating blue star of spectral type B7V with a visual magnitude of 1.35\,mag, was used to correct for these contaminants. This telluric spectrum, which has fewer resolved stellar lines, effectively represents the atmospheric and interstellar medium spectra. The Lunar limb and disk spectra were corrected by subtracting and scaling HR3982's spectrum using IRAF's TELLURIC task. This correction process removed telluric and interstellar features, ensuring accurate analysis of the lunar spectra.

\subsubsection{Solar scatter component}
Solar light scattered from the lunar surface contaminates the observed exospheric spectra. A series of procedures were implemented to correct this. First, multiple spectra of the lunar disk, representing the scattered light component, were acquired. Each disk spectrum was normalized and aligned using cross-correlation and dispersion axis shifting. A combined disk spectrum, created by median combining these spectra with the IRAF task `scombine', represented the solar scatter component. \cite{contarini1996} also adopts a similar methodology in their analysis. The lunar limb spectrum, including the exosphere and disk contamination, was similarly corrected for sub-pixel shifts. To match the spectral resolution, the limb spectrum was matched to the median disk spectrum using Spectral Resolution Matching from the SDSS MaNGA Data Analysis Pipeline \citep{westfall2019, belfiore2019}. After aligning the limb spectra and matching the resolution, the lunar exosphere spectrum was extracted by dividing the median disk spectrum from the limb spectrum. The resulting exospheric spectra reveal the Na D1 and D2 lines at 5895.93\,$\textrm{\AA}$ and 5889.96\,$\textrm{\AA}$, respectively, with an average SNR of $\sim$128, as shown in Figure \ref{fig:NaExosphere}. To demonstrate the efficiency of the disk removal of our method, we also show the Fe\,I (5883.82\,\text{\AA}) and Ni\,I (5892.88\,\text{\AA}) absorption features in the limb and disk region in Figure \ref{fig:NaExosphere}. The exospheric profile for both lines shows no emission feature, with the intensity remaining constant at around 1.0. Therefore, the dividing out process efficiently removed the disk component from the limb spectra.\\
\\
\subsection{Flux calibration}
Flux calibration of the data was performed using the standard star HD17573, which was observed during our observations. HD17573 has a spectral type of B8Vn, with an effective temperature of approximately 13000\,K. The star is located at a distance of 48.9 pc and has a radius of 4.51 $R_{\odot}$. A high-resolution synthetic model of the star was generated using the Coelho Synthetic Stellar Library \citep{coelho2014}. We assumed an [Fe/H] of 0.2, a log(g) value of 4, and a solar-like value for the alpha elements over iron abundance. The generated high-resolution spectrum was binned to match the resolution of the observed spectrum. The continuum of both spectra was modelled with smooth univariate spline functions. The ratio of the continuum provided the response curve, which was subsequently used for flux calibrating the observed exospheric line strength measurements in units of KiloRayleighs (kR). The calibrated flux measurements are given in Table \ref{table:flux_measurements}.


\section{Results}\label{sec:results}

\begin{figure*}
    \centering
    \includegraphics[width=0.32\textwidth]{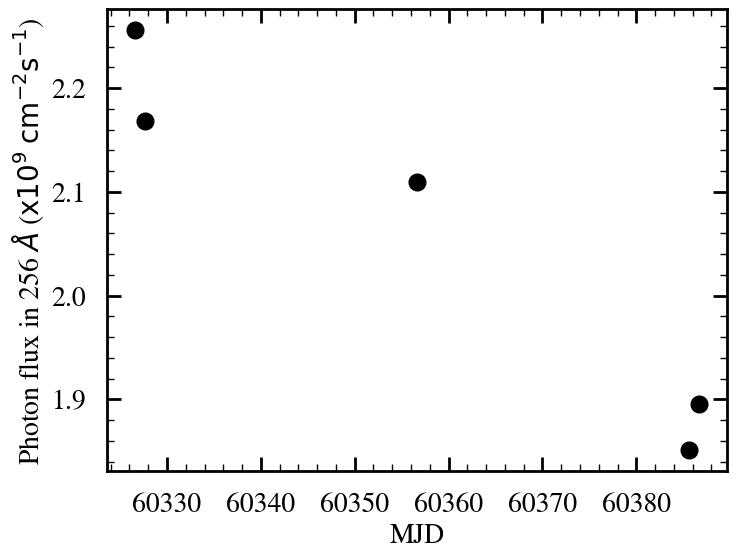}
    \includegraphics[width=0.32\textwidth]{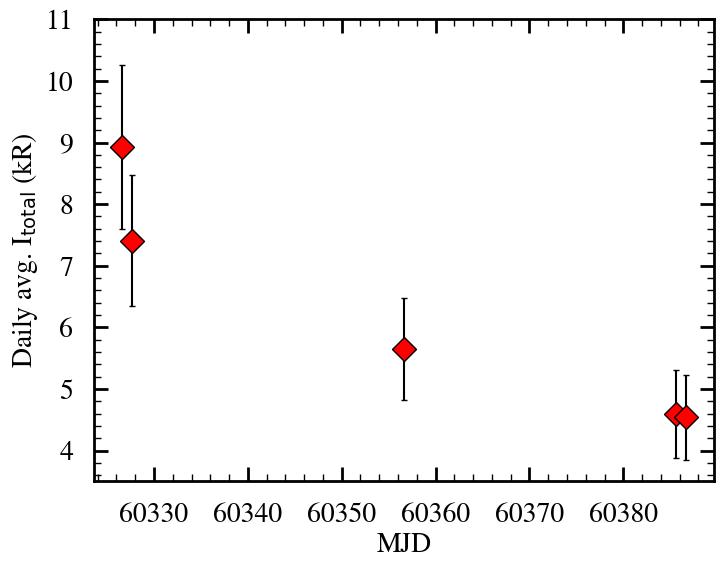}
    \includegraphics[width=0.32\textwidth]{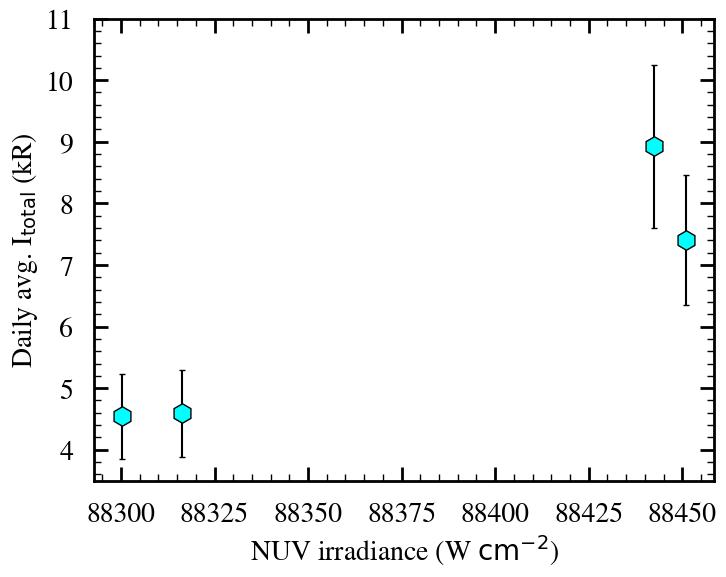}
    
   \caption{The left panel presents the decrease in EUV photon flux (256\,\text{\AA}) over MJDs observed, and the middle panel shows the corresponding declining trend we observe in the daily averaged $I_{Total}$ over the same period. An increasing trend is observed between the daily averaged integrated flux from NUV (2000 - 4000\,\text{\AA}) and daily averaged $I_{Total}$, in the right panel. Note that the NUV flux from TSIS-1 was not available on 16 February 2024.}
    \label{fig:GOES-VBT_comparison}
\end{figure*}

The lunar exospheric spectrum was obtained by dividing the observed median disk spectrum from the limb spectrum. Figure \ref{fig:NaExosphere} shows the emission peaks of Na\,I D2 (5889.95\,\text{\AA}) and Na\,I D1 (5895.92\,\text{\AA}) detected in the lunar exosphere. Lunar minerals such as ilmenite ($FeTiO_{3}$), olivine (($Mg, Fe)_{2}$$ SiO_{4}$), and pyroxene contain Fe, a refractory element. However, the exosphere lacks significant Fe\,I emission, suggesting that iron is less abundant than sodium in the exosphere, likely due to its lower cross-section in the release processes \citep{flynn_stern1996, sarantos2012metallic}.

The Na\,I D2 flux observed in this study corresponds to angular heights below $\sim$590 km. The line intensities above the continuum of the Na lines were determined by fitting Gaussian profiles to the exospheric spectra using SPLOT in IRAF. These intensities measured and calibrated in kR units are shown in Table \ref{table:flux_measurements}. Fluxes ranging from 6.13\,kR to 2.8\,kR for the D2 line and from 3.5\,kR to 1.3\,kR for the D1 line were observed. These measurements, taken within the lunar phase angle range of 83 to 96 degrees, are consistent with previous estimates in literature \citep{potter1988a, Potter1988b, sprague1992, potter_and_morgan1991, potter_and_morgan1994}. The theoretical line flux ratio of Na\,I D2 to Na\,I D1 is estimated to be 2.0 \citep{sansonetti2008}. In our observations, the average line flux ratio is 1.79$\pm$0.14, with a maximum of 2.46$\pm$0.69, closely matching the theoretical value and aligning with the previously estimated range of 1.23 to 2.47 by \cite{potter1988a} and \cite{berezhnoy2014}.

\subsection{EUV correlation}
The simultaneous $F_{GOES}$ from GOES-EUVS was obtained. The Photon flux in $cm^{-2}s^{-1}$ for six spectral lines are provided in Table \ref{table:flux_measurements}. The spectral lines GOES gives are 256, 284, 304, 1175, 1335, and 1405\,\text{\AA}. Among these, the first three bands are considered as EUV, and the last three are considered as FUV.  

Figure \ref{fig:GOES-VBT_comparison} presents the variation of $F_{GOES}$ and total optical line flux ($I_{Total}$, the sum of the measured intensities of Na\,I D2 and D1 lines) with respect to time, along with the relationship between daily averaged NUV irradiance from SIM. The daily averaged $I_{Total}$ here is the average of the sum of $I_{Total}$ each day of observation. The left panel of Figure \ref{fig:GOES-VBT_comparison} show the temporal variation in the photon flux from the 256\,\text{\AA} EUV band. During our observation period, a decrease in the EUV band flux from about $2.15\times10^{9}$\,$cm^{-2}s^{-1}$ to approximately $1.85\times10^{9}$\,$cm^{-2}s^{-1}$ is observed. Correspondingly, the daily averaged $I_{Total}$ also decreases during the same period, with a significant decline from $\sim$9\,kR at MJD 60330 to around 5\,kR at MJD 60380.

\begin{figure}
\centering
\includegraphics[width=\columnwidth]{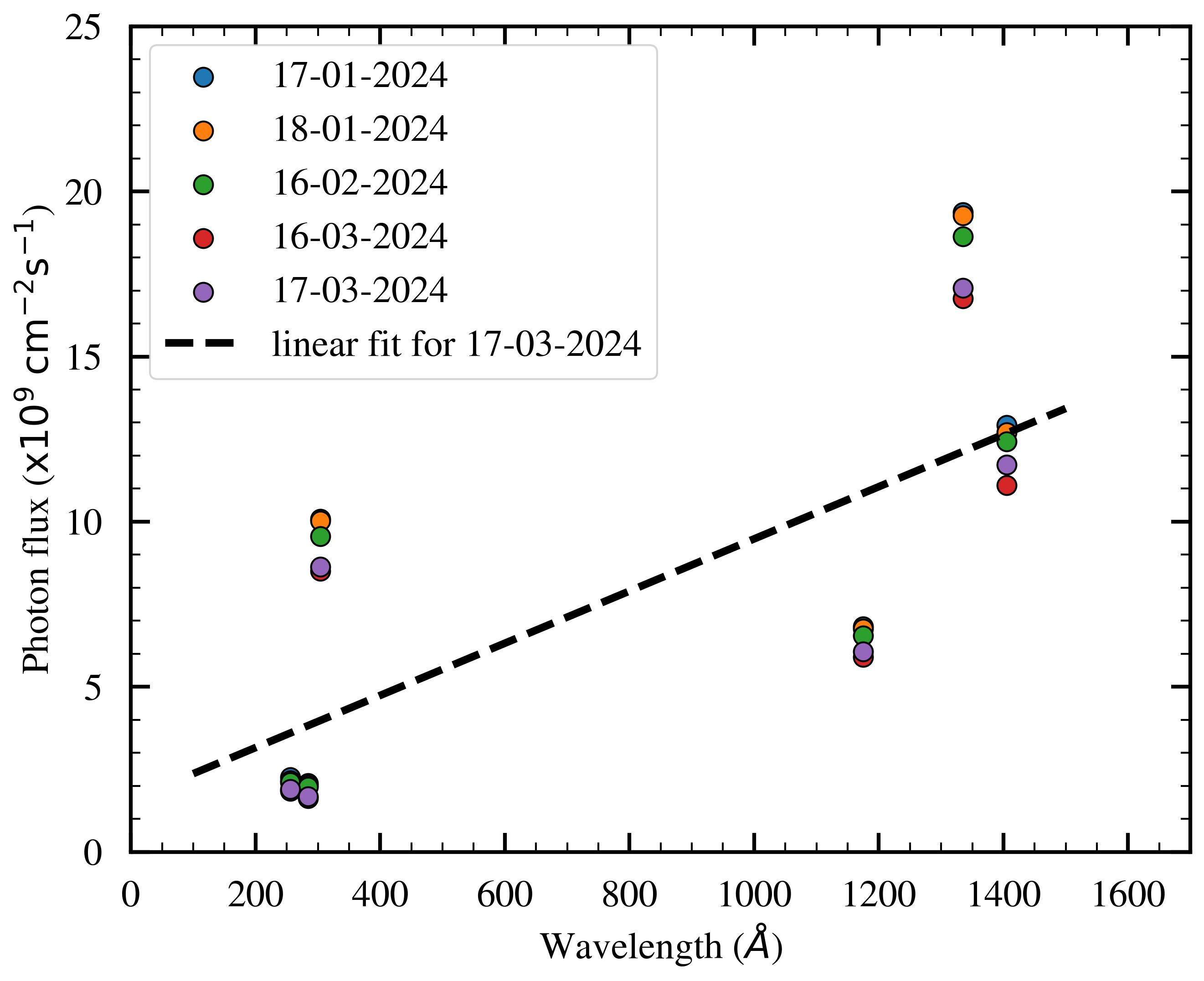}
\caption{The figure illustrates the photon flux reaching the Moon across different wavelengths. The flux decreases with increasing UV photon energy, with FUV exhibiting higher flux compared to EUV. This trend is consistent across all observation nights. A linear fit (dashed line) is shown for the data from 17-03-2024 to highlight the decreasing trend.}
\label{fig:Photonflux_vs_wavelength}
\end{figure}

The Pearson and Spearman correlation tests were performed on the correlation of daily averaged $I_{total}$ and photon flux in $F_{GOES}$. The Pearson test measures the linear relationship, assuming a linear correlation between the datasets. In contrast, the Spearman test evaluates the strength and direction of a monotonic relationship, making it suitable for both linear and non-linear data. The Pearson test yielded a p-value of $3.31\times10^{-4}$ and a correlation coefficient of 0.834, while the Spearman test resulted in a p-value of $5.28\times10^{-6}$ and a correlation coefficient of 0.956. These values indicate a strong, statistically significant positive correlation, suggesting the possible presence of more than one component in the relationship. Additionally, we obtained daily averaged NUV irradiance from TSIS-1 for each observation day, except for 16 February 2024. An increase in the daily averaged $I_{Total}$ was observed with a corresponding rise in NUV irradiance.

Our analysis checked the relation between the photon flux in all the EUVS wavelengths from GOES. It was found that the Na emission from the Lunar exosphere increases with an increase in the EUVS photon flux reaching the lunar surface. It is also seen that the increase is nonlinear. To check how each correlation varies with EUVS wavelengths, we fitted a non-linear mathematical function given in Equation \ref{eqn:fitted_nonlinear_function} to each correlation. 
\begin{equation}
\label{eqn:fitted_nonlinear_function}
    \Phi_{Na} = A \times \Phi_{EUV}^{\alpha} + B
\end{equation}

\begin{figure*}
    \centering
    \begin{subfigure}[b]{0.33\textwidth}
        \centering
        \includegraphics[width=\textwidth]{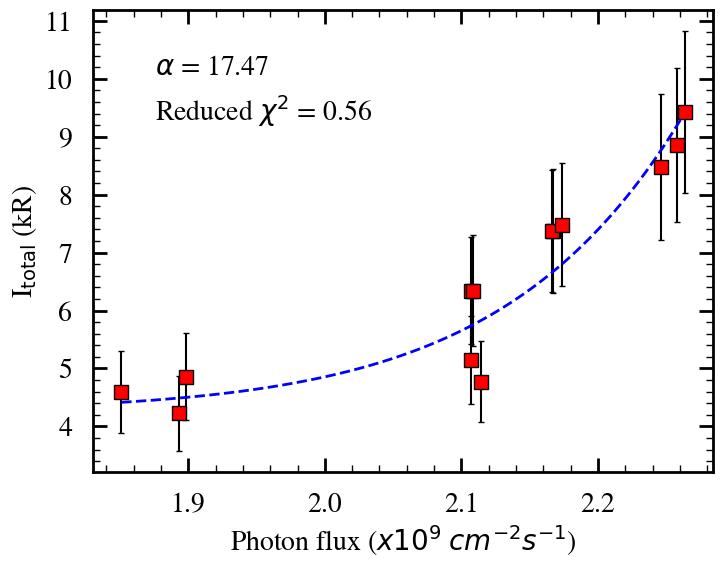}
        \caption{256\,\text{\AA}}
        \label{fig:256}
    \end{subfigure}
    \hfill
    \begin{subfigure}[b]{0.33\textwidth}
        \centering
        \includegraphics[width=\textwidth]{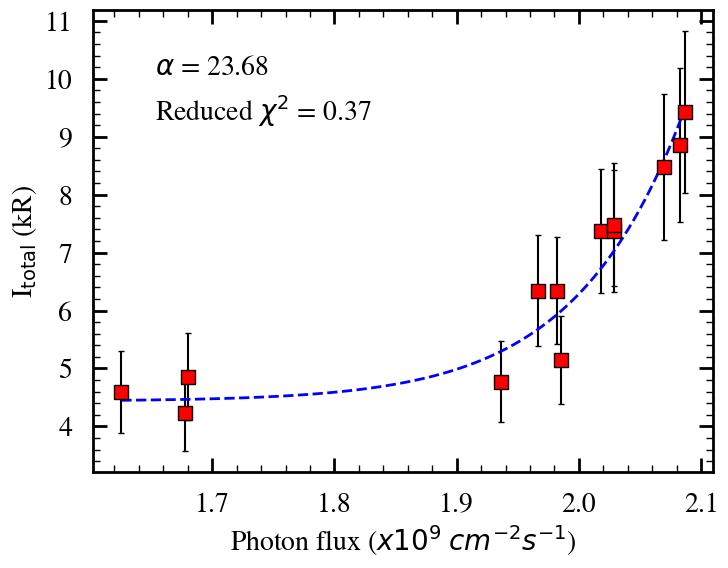}
        \caption{284\,\text{\AA}}
        \label{fig:284}
    \end{subfigure}
    \hfill
    \begin{subfigure}[b]{0.33\textwidth}
        \centering
        \includegraphics[width=\textwidth]{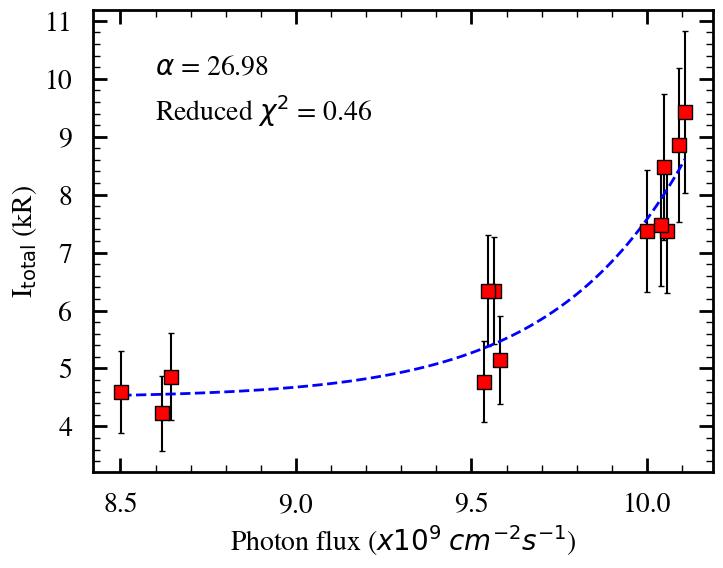}
        \caption{304\,\text{\AA}} 
        \label{fig:304}
    \end{subfigure}
    
    \vspace{1em}
    
    \begin{subfigure}[b]{0.33\textwidth}
        \centering
        \includegraphics[width=\textwidth]{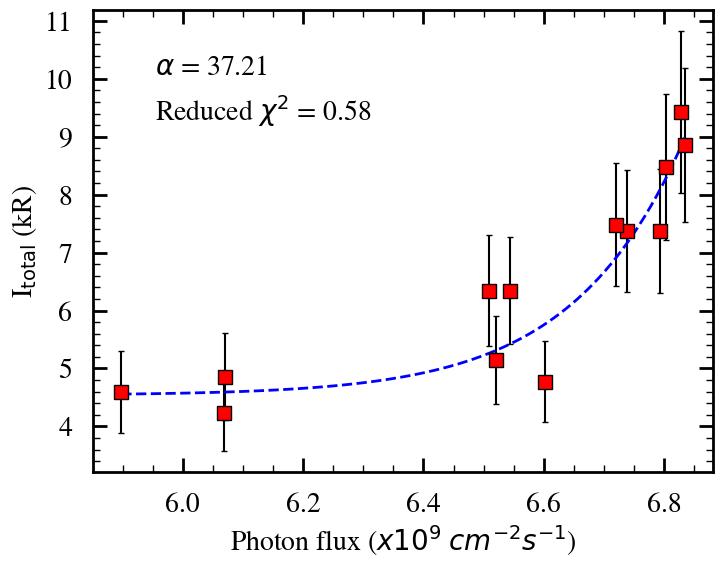}
        \caption{1175\,\text{\AA}}
        \label{fig:1175}
    \end{subfigure}
    \hfill
    \begin{subfigure}[b]{0.33\textwidth}
        \centering
        \includegraphics[width=\textwidth]{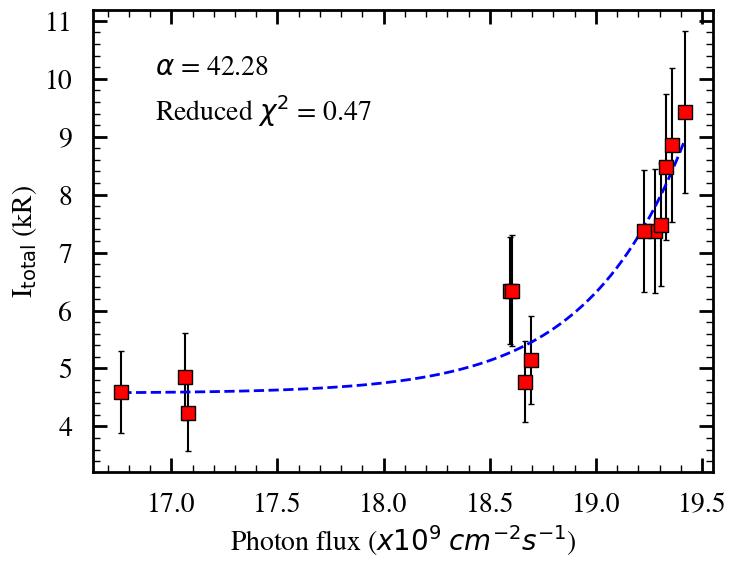}
        \caption{1335\,\text{\AA}}
        \label{fig:1335}
    \end{subfigure}
    \hfill
    \begin{subfigure}[b]{0.33\textwidth}
        \centering
        \includegraphics[width=\textwidth]{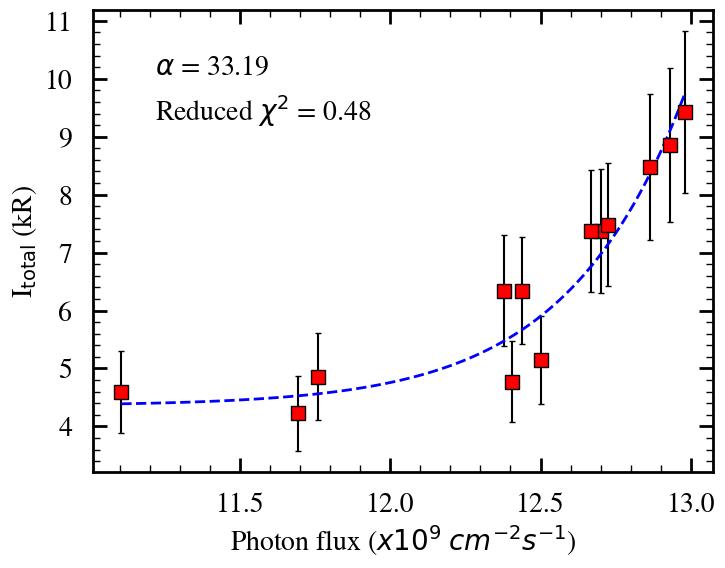}
        \caption{1405\,\text{\AA}}
        \label{fig:1405}
    \end{subfigure}

    \caption{The figure illustrates the relationship between the EUV photon flux obtained from GOES and the $I_{Total}$ of Na\,I D2 + D1 lines in the lunar exosphere. Panels (a) to (c) depict this correlation for EUV photon fluxes at wavelengths of 256\,\text{\AA}, 284\,\text{\AA}, and 304\,\text{\AA}, respectively. Panels (d) to (f) show the correlation of FUV photon fluxes at 1175\,\text{\AA}, 1335\,\text{\AA}, and 1405\,\text{\AA}, respectively. The X-axis represents the photon flux in units of $\times10^{9}$\,$cm^{-2}s^{-1}$, while the Y-axis indicates the $I_{Total}$ in kR. The blue dashed lines show the fitted curves obtained using Equation \ref{eqn:fitted_nonlinear_function}. These fits reveal a nonlinear relationship, indicating that higher photon fluxes correspond to an increase in the total Na\,I D2 + D1 flux.}
    \label{fig:EUV_photonflux_Optical}
\end{figure*}

In Equation \ref{eqn:fitted_nonlinear_function}, $\Phi_{Na}$ is the total Na flux is kR, $\Phi_{EUV}$ is the photon flux, $\alpha$ is the growth factor, A and B are constants. The growth factor estimated by fitting Equation \ref{eqn:fitted_nonlinear_function} at different EUVS wavelengths are plotted in Figure \ref{fig:EUV_photonflux_Optical}. The FUV wavelengths exhibit a higher average growth factor compared to the EUV wavelengths, suggesting increased non-linearity in the relationship between $I_{Total}$ of Na and the FUV photon flux. The decreasing trend in growth factor from FUV to EUV wavelengths aligns with the decline in photon flux across these wavelengths, as shown in Figure \ref{fig:Photonflux_vs_wavelength}.

The flux of Na released from a silicate-based substrate is expected to be linearly proportional to the EUV photon flux \citep{Yakshinskiy1999, Wurz2022}. Hence, a higher deviation from linearity indicates a greater contribution from other non-thermal processes, such as sputtering, and a reduced role of PSD. The EUV wavelengths show a higher contribution from PSD, possibly indicating that the efficiency of the PSD process increases at these wavelengths.

\begin{table*}
\caption{The $T_{ch}$ and scale heights were derived from the Doppler broadening of the Na\,I D2 line in the lunar exosphere spectrum. The observed broadening ($\Delta\lambda_{obs}$) was corrected for instrumental effects to obtain the Doppler broadening. The heliocentric radial velocity ($RV_{hc}$) of the Moon during observations, the corresponding g-value estimated from $RV_{hc}$, and the sodium $N_{los}$ and $N_{zen}$ estimations are also presented.}

\label{table:CT_and_SH_measurements}  
\centering  
\begin{tabular}{lccccccc}
\hline\hline 
\noalign{\smallskip}
MJD              & \begin{tabular}[c]{@{}c@{}}$\Delta\lambda_{obs}$\\ (\text{\AA})\end{tabular} & \begin{tabular}[c]{@{}c@{}}$T_{ch}$ \\ (K)\end{tabular} & \begin{tabular}[c]{@{}c@{}}Scale Height \\(km)\end{tabular} & \begin{tabular}[c]{@{}c@{}}$RV_{hc}$ \\( $km\,s^{-1}$)\end{tabular} & \begin{tabular}[c]{@{}c@{}}g-value\\($photons\,s^{-1}\,atom^{-1}$)\end{tabular} & \begin{tabular}[c]{@{}c@{}}$N_{los}$\\($\times10^{9}$ $atoms\,cm^{-2}$)\end{tabular} & \begin{tabular}[c]{@{}c@{}}$N_{zen}$\\($\times10^{9}$ $atoms\,cm^{-2}$)\end{tabular}\\ \noalign{\smallskip} \hline
\noalign{\smallskip}
60326.607 & 0.108 & 6195 & 1377 & 1.051 & 0.74 & 8.2$\pm$1.7 & 4.9$\pm$1.0 \\
60326.621 & 0.104 & 4981 & 1107 & 1.051 & 0.72 & 8.2$\pm$1.7 & 4.9$\pm$1.0 \\
60326.649 & 0.107 & 5778 & 1285 & 1.051 & 0.73 & 7.5$\pm$1.5 & 4.5$\pm$0.9 \\
\noalign{\smallskip}
\noalign{\smallskip}
60327.568 & 0.112 & 7425 & 1651 & 1.040 & 0.77 & 5.8$\pm$1.2 & 3.5$\pm$0.7 \\
60327.657 & 0.108 & 6290 & 1398 & 1.040 & 0.74 & 5.6$\pm$1.1 & 3.3$\pm$0.7 \\
60327.672 & 0.112 & 7196 & 1600 & 1.039 & 0.77 & 5.2$\pm$1.1 & 3.1$\pm$0.6 \\
\noalign{\smallskip}
\noalign{\smallskip}
60356.578 & 0.099 & 3717 & 826 & 1.029 & 0.68 & 4.8$\pm$1.0 & 2.8$\pm$0.6 \\
60356.588 & 0.098 & 3413 & 759 & 1.029 & 0.68 & 5.5$\pm$1.1 & 3.2$\pm$0.7 \\
60356.614 & 0.111 & 6983 & 1552 & 1.029 & 0.76 & 5.6$\pm$1.2 & 3.4$\pm$0.7 \\
60356.661 & 0.103 & 4850 & 1078 & 1.028 & 0.71 & 4.1$\pm$0.8 & 2.4$\pm$0.5 \\
\noalign{\smallskip}
\noalign{\smallskip}
60385.586 & 0.12 & 9789 & 2176 & 1.018 & 0.82 & 4.0$\pm$0.8 & 2.4$\pm$0.5 \\
\noalign{\smallskip}
\noalign{\smallskip}
60386.661 & 0.106 & 5558 & 1236 & 1.001 & 0.72 & 4.8$\pm$1.0 & 2.9$\pm$0.6 \\
60386.693 & 0.136 & 15198 & 3379 & 1.001 & 0.94 & 3.0$\pm$0.6 & 1.8$\pm$0.4 \\ \hline
\end{tabular}
\end{table*}

\subsection{Column Density}
\begin{figure}
    \centering
    \includegraphics[width=\columnwidth]{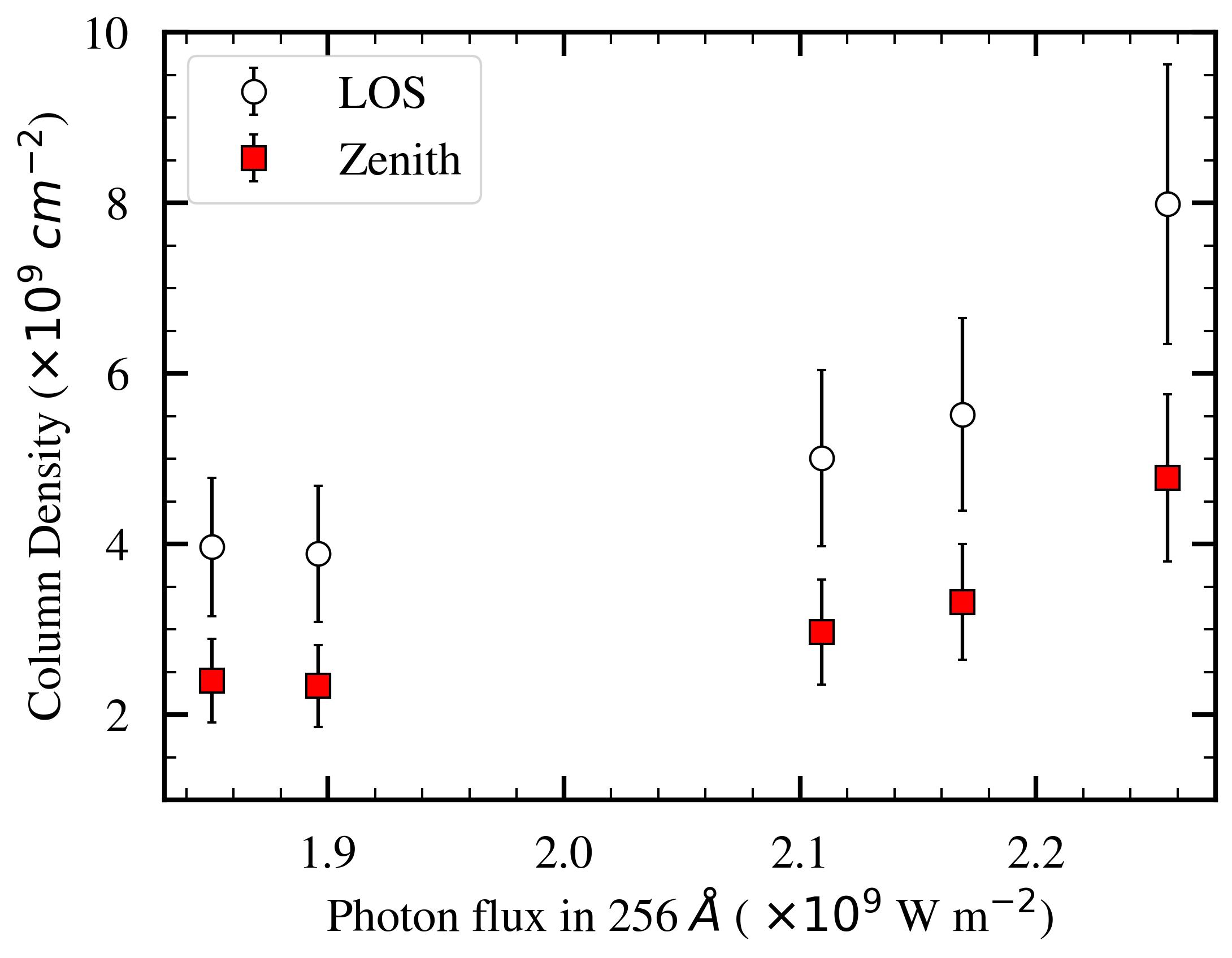}
    \caption{The non-linear variation of day-wise averaged $N_{los}$ and $N_{zen}$  with respect to the EUV photon flux in the 256\,\text{\AA} wavelength. The $N_{los}$ is corrected for the variation in g-value and Sun-Moon distance variation.}
    \label{fig:LOS_Ncol_PhotonFlux}
\end{figure}

Although the Sun-Moon distance correction was applied to both $F_{GOES}$ and NUV irradiance, the changing g-values (the number of photons emitted per second per Na atom) may still introduce a dependency on $I_{Total}$ due to variation of -1.5 and +1.5\,$km\,s^{-1}$ in the heliocentric radial velocity ($RV_{hc}$) during lunar orbital motion \citep{berezhnoi2023}. The g-value is also highly sensitive to the solar spectrum shift and the temperature of exospheric atoms. To address the dependence on varying g-value during our observations, we estimated the sodium column density along the line of sight ($N_{los}$, $atoms\, cm^{-2}$) from the observed Na\,I D2 line intensity in Rayleighs \citep{potter2000, berezhnoi2023} by applying the variable g-value corresponding to the $RV_{hc}$ for the observation period in Equation \ref{equ:numberdensity}.
\begin{equation}\label{equ:numberdensity}
N_{los} = 10^{9} \times \frac{I_{obs}}{g}
\end{equation}
In Equation \ref{equ:numberdensity}, $I_{obs}$ represents the observed brightness of the Na\,I D2 line in kR, and $g$ is the g-value. \cite{berezhnoy2014}, and several previous studies have typically assumed the g-value to be constant $\sim$0.53 $photons\,s^{-1}\,atom^{-1}$ when estimating $N_{los}$ \citep{potter2000, killen2019, sarantos2010}. However, \cite{berezhnoi2023} quotes that the g-values (at 0 $km\,s^{-1}$ heliocentric velocity) for the Na D2 line employed in various studies differ by approximately 10\%. \cite{berezhnoi2023} also estimates g-value variations, assuming the lunar exosphere to be optically thin, collisionless, and without accounting for atomic stimulated emission. From Figure 3 of \cite{berezhnoi2023}, we estimated the g-values for temperatures ranging from 500\,K to 5500\,K at an $RV_{hc}$, which exhibited a linear correlation. A linear fit was applied to this data to estimate the g-values corresponding to the observed exospheric temperature. The estimated $RV_{hc}$, g-value, and $N_{los}$ are given in Table \ref{table:CT_and_SH_measurements}. A variation in the Na\,I D2 line g-values, ranging from 0.676 to 0.941, is estimated. This variation in g-values across observations was accounted for while estimating the $N_{los}$. The results indicate an average $N_{los}$ of $5.6\times10^{9}$$atoms\,cm^{-2}$. In comparison, \cite{mendillo1999} reported a near-surface line-of-sight neutral sodium column density of $1.4\times10^{9}$ $atoms\,cm^{-2}$, which is lower than the value derived in this study. We also estimate the zenith column density ($N_{zen}$), with the formulation adopted from \cite{berezhnoy2014} and \cite{Chamberlain1963}. An average $N_{zen}$ of $3.3\times10^{9}$$atoms\,cm^{-2}$ was estimated. The $N_{zen}$ varied from $1.8\times10^{9}$ to $4.9\times10^{9}$ $atoms\,cm^{-2}$ within a phase angle variation of 83 to 96 degrees (lunar illumination of $\sim$43\% to 55\%). The variation in phase angle throughout the observation period corresponds to a change in $N_{zen}$ of $7.9 \times 10^{8}$ $atoms\ cm^{-2}$. These values are also given in Table \ref{table:CT_and_SH_measurements}. The daily averaged $N_{los}$ and $N_{zen}$ were plotted against daily averaged $F_{GOES}$ at 256\,\text{\AA} corrected for variation in sun-moon distance. The non-linear correlation seen in Figure \ref{fig:EUV_photonflux_Optical} remains evident even after applying the corrections, as demonstrated in Figure~\ref{fig:LOS_Ncol_PhotonFlux}.

We examined Fe I in the spectrum to estimate the lower limit of its $N_{zen}$ in the lunar exosphere. Fe I, being a refractory element, is a potential exospheric species \citep{sarantos2012} with a g-value of 0.00457 photons atom$^{-1}$ s$^{-1}$ \citep{sarantos2012metallic}. Assuming a 5-sigma detection threshold, the estimated $N_{zen}$ for Fe I is $3.9 \times 10^{10}$ atoms cm$^{-2}$, consistent with earlier estimates of $1.9 \times 10^{10}$ and $2.5 \times 10^{10}$ atoms cm$^{-2}$ by \cite{berezhnoy2014} and \cite{flynn_stern1996}, respectively. 

\subsection{Characteristic Temperature and Scale Height}
We also estimated the Characteristic Temperature ($T_{ch}$) and Scale height. $T_{ch}$, also known as the exospheric temperature, describes the kinetic temperature of particles in the exosphere of the Moon. $T_{ch}$ is critical in understanding the thermal and dynamic properties of the exosphere, where collisions between particles become rare and ballistic trajectories dominate, which is the case of the Lunar exosphere \citep{Killen2018, Leblanc2022}. The primary cause of line broadening in the exosphere is the Doppler effect, which is due to the thermal motion of the particles. The relationship between the Doppler broadening in \text{\AA} ($\Delta\lambda_{D}$) of the Na line and $T_{ch}$ in Kelvin is given in Equation \ref{eqn:characteristic_temperature}.
\begin{equation} 
\label{eqn:characteristic_temperature}
    T_{ch} = \frac{m_{Na} c^{2} \Delta\lambda_{D}^{2}}{8k\lambda_{o}^{2}}
\end{equation}
The constants in Equation \ref{eqn:characteristic_temperature} are $m_{Na}$, the mass of a neutral Na atom, which is taken as $3.82 \times 10^{-26}$\,kg; $c$, the speed of light; and $k$, the Boltzmann constant. The Doppler broadening was estimated from the Na\,I D2 lines in our observations after accounting for the instrumental effects. The estimated average Doppler broadening was 0.07\,\text{\AA}. These broadening measurements were applied in Equation \ref{eqn:characteristic_temperature} to estimate $T_{ch}$. The average $T_{ch}$ was estimated to be 6721\,K, with most estimates falling within the range of 3400\,K to 8000\,K. The scale heights for all $T_{ch}$ were also estimated, and the average scale height was 1494\,km. The scale heights estimated in our analysis range between 700\,km to 1700\,km. However, two measurements on 16 March 2024 and 17 March 2024 indicated temperatures of approximately 9200\,K and 15,200\,K, respectively, with corresponding scale heights of $\sim$2200\,km and $\sim$3400\,km. The observed linewidth, $T_{ch}$, and scale heights are given in Table \ref{table:CT_and_SH_measurements}. The $\Delta\lambda_{obs}$ was obtained by fitting a Gaussian profile to the Na\,I D2 line. The mean percentage error in fitting is $\sim$1.2\%. The Doppler broadening was estimated by de-convolving the instrumental broadening from $\Delta\lambda_{obs}$. The mean instrumental broadening during our nights was estimated to be 0.083\,\text{\AA}, with an uncertainty of 7.2\%. The temperatures derived from these estimates had a mean percentage error of 23.2\%. The average uncertainty percentage in g-values is 4.2\%.

Data from the Solar Ultraviolet Imager (SUVI) onboard GOES list solar flares, including their UTC timestamps and flare classes \citep{Darnel2022}. The data from SUVI indicates solar flares on 2024-03-16 14:16:00, 2024-03-17 15:43:00, 2024-03-17 15:58:00, and 2024-03-17 15:58:00, with flare classes C1.6, C4.9, C3.3, and C3.3, respectively. These solar flares occurred very close in time to our observations on 16 (MJD 60385.586) and 17 March 2024 (MJD 60386.693). \cite{Smith1974} and \cite{Bruevich2017} noted that during periods of high solar activity, particularly major flares, the solar flux in the X-ray and EUV regions can vary by several orders of magnitude. \cite{Bruevich2017} further observed based on GOES-15 data from Solar Cycle 24 that even minor variations in solar flux, whether due to solar cycle activity or large flares, can significantly impact UV and X-ray activity indices. Additionally, radiative hydrodynamic simulations by \cite{Allred2005} for moderate and strong solar flares show increased coronal and transition region densities, leading to substantial enhancements in both UV and optical line and continuum emissions. Therefore, solar flares occurred close to our observation period could cause increase the solar UV emission, possibly explaining the higher observed $T_{ch}$ and scale height. In addition, \cite{kuruppuaratchi2023} observed that temperatures derived from linewidth measurements were consistently lower than those obtained from coronagraphic intensity altitude profiles. Therefore, the $T_{ch}$ values from our analysis may be underestimated.


\begin{figure*}
    \centering
    \includegraphics[width=2\columnwidth]{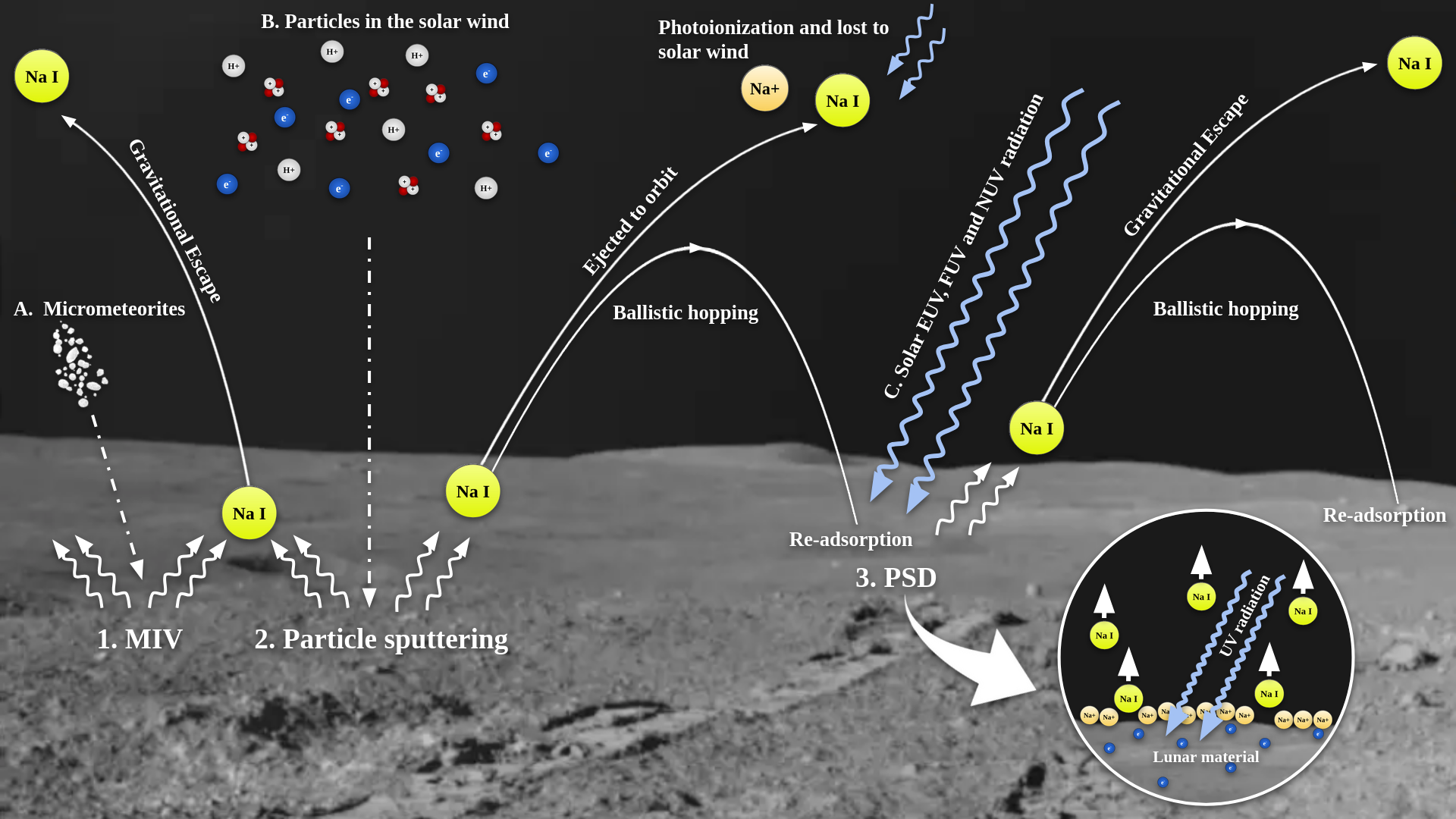}
    \caption{The illustration shows the source, sink, and non-thermal release processes involved in maintaining the lunar exosphere. The major sources contributing to non-thermal release processes are (A) Micrometeorites, (B) Particles in the solar wind, and (C) EUV, FUV, and NUV radiation of solar origin. These sources cause (1) Micrometeorite Impact Vaporization (MIV), (2) Particle Sputtering, and (3) Photon Stimulated Desorption (PSD). Neutral Na atoms released through these processes can undergo gravitational escape if sufficient energy is imparted by sources A, B, or C. These atoms may become part of the Na tail and be lost to the interplanetary medium. A fraction of neutral Na remains in orbit around the Moon in the exosphere or follows a ballistic trajectory. The atoms fall back to the surface and get reabsorbed, forming an adsorbate layer. Additionally, neutral Na in the exosphere can be photoionized by UV radiation from the sun and lost to space, contributing to the Moon's Na tail. The inset in the illustration demonstrates the PSD process. Na is adsorbed onto lunar material as Na ions, held by electrostatic forces in an adsorbate layer. UV radiation from the sun is absorbed by the lunar bulk material, creating free electrons. These free electrons combine with Na ions to form neutral Na. The size of the neutral atom ($2.8\,\text{\AA}$) is larger than the ion ($1.4\,\text{\AA}$), causing repulsive motion and releasing neutral Na into the exosphere. The background image used in this is from the Pragyan rover in Chandrayaan-3 (credits: ISRO).}
    \label{fig:illustration}
\end{figure*}

\section{Discussion}\label{sec:descussion}
The Na abundance in the lunar regolith is primarily found in minerals such as sodium orthosilicate, sodium metasilicate, and albite \citep{heiken1991}. Furthermore, \cite{Narendranath2022} suggest that in addition to the geochemical fraction, the distribution of Na in the upper layers of the regolith also includes an adsorbate component. A significant fraction of sodium atoms may form weak bonds with metal and metal-oxide surfaces. These bonds can be broken by UV photons, causing the atoms to desorb and return to the exosphere. Hence, from the existing understanding, the Na population in the exosphere results from the release of Na from the adsorbate layer through PSD and Na-bearing lunar surface minerals through other processes \citep{heiken1991, Colaprete2016}.

The release of Na from the lunar surface to the exosphere occurs through four main processes: thermal desorption, PSD, sputtering by solar wind ions, and MIV. Thermal desorption, a thermal process, involves particles in thermal equilibrium with the lunar surface, usually in the $T_{ch}$ range of 100 to 400\,K \citep{Dukes2017}. In contrast, PSD, sputtering, and MIV are classified as supra-thermal processes due to the higher energies involved. These processes are expected to have $T_{ch}$ exceeding 1200\,K \citep{Yakshinskiy1999, yakshinskiy2000, yakshinskiy2004, sarantos2010}.

Supra-thermal processes involve high-energy photons or particles, which are driven by energetic interactions. Agents such as micrometeorites and solar particles impart energy to the lunar surface through momentum transfer. This energy exchange results in the release of Na into the exosphere. Among the processes, PSD plays a significant role in the release of Na atoms from the lunar regolith \citep{Nicole2024}. High-energy UV from the Sun provides enough energy to overcome the binding energy of Na atoms, ejecting them into the exosphere with non-Maxwellian velocity distributions. Unlike thermal processes or momentum transfer, PSD is explained to be driven by electronic excitations of the lunar surface and releasing the Na adsorbate layer. When solar UV photons strike the surface, substrate-mediated excitations occur, transferring energy to an electron in the solid’s valence band, exciting it to the conduction band. Alkali metals like Na or K, typically adsorbed in an ionic state, undergo desorption as the excited electron transfers to the adsorbed $Na^{+}$ or $K^{+}$. The neutralized Na or K, having a larger atomic radius than the ion, is ejected from the surface due to repulsion, leading to desorption \citep{Dukes2017}. Laboratory experiments on a model $SiO_{2}$ substrate suggest that PSD can be a primary mechanism for Na desorption from the Moon. Experiments demonstrated that UV photons efficiently desorb Na atoms at surface temperatures around 250\,K \citep{Mandey1998, Yakshinskiy1999, yakshinskiy2000}. Further measurements using a lunar basalt sample, a more complex oxide substrate, confirmed that PSD of Na is indeed an efficient desorption process on the Moon \citep{yakshinskiy2004}. 

Although PSD is an efficient process, the gravitationally escaping and higher-altitude Na population has a lesser contribution from PSD. \cite{yakshinskiy2004} noted that UV solar photons, responsible for PSD, do not significantly contribute to the Na flux with speeds greater than 2\,$km\,s^{-1}$. This suggests that PSD is less effective in releasing Na atoms to higher altitudes or even escaping Lunar gravity \citep{Vervack2010, Mouawad2011, Tenishev2013}. Some studies also indicate that PSD does not play a major role in releasing Na atoms bound in primary minerals \citep{Nicole2024, sprague1992, Wurz2022}, as it lacks the energy to overcome the lattice binding energies of lunar minerals. Therefore, a more energetic process is required to release sodium to the higher altitudes observed in past studies \citep{Leblanc2008}. 

The EUVS bands from GOES, ranging from 256\,\text{\AA} to 1405\,\text{\AA}, considered in our analysis have photon energies between 8.8 - 48.5\,eV, which is higher than those used in the previous studies. Since EUV and FUV could possibly impart higher energies, we propose that PSD caused by these radiations could also contribute to the high altitude abundance of Na by releasing the Na adsorbate layer and Na bound in a mineral. Given the recent variations in solar activity and its overlap with our observation period, we noted a gradual decline in EUVS flux over three months measured using GOES (see Figure \ref{fig:GOES-VBT_comparison}). We see a strong correlation between the Na abundance in the Lunar exosphere during our epoch of observation and the EUVS flux greater than 8.8\,eV. This indicates that the lunar surface material can effectively release the Na via PSD by the UV photons with energies in the FUV and EUV regime from the Sun, thereby contributing to the Na exosphere. 

The binding energy or the cohesive energy values for Na found in lunar plagioclase feldspar range from 1.1 to 2.65\,eV \citep{kittel2021, lammer2003, mcgrath1986, cheng1987}. \cite{sprague1992} mentions that desorption of Na from the surface and through pores on the surface is explained by photo-desorption by the photons of energy range 1.5 to 4\,eV (8000 - 3000\,\text{\AA}). \cite{Morrissey2022} utilize Molecular dynamics calculations to accurately measure the binding energy of various Lunar minerals like plagioclase feldspars and sodium pyroxenes. They indicate that minerals like Sodium orthosilicate, Sodium metasilicate, and Albite have binding energies of 2.6\,eV, 4.4\,eV, and 7.9\,eV, respectively. In addition, they note that the binding energy of Na depends on the substrate considered, and all the minerals considered contain Na–O bonds have different binding energies. This is the energy required to remove Na atoms from the mineral structure, reflecting the strength of the chemical bonds and also causing the release of the adsorbate layer. The EUVS wavelengths considered in this work have photon energy ranging between 8.8\,eV and 48.4\,eV. This radiation could possibly impart space weathering by removing Na bounded in minerals by breaking bonds as well as releasing the Na adsorbate layer. Hence, this work is observationally indicating that the lunar surface material can effectively release the Na held as an adsorbate layer or bounded in a mineral via PSD by the UV photons with energies in the EUV regime from the Sun, thereby contributing to the Na exosphere.

In addition, the atomic metal abundances in the exosphere are expected to reflect the surface abundances of these metals \citep{taylor1982, johnson1991}. Initially, simple stoichiometric models predicted that major elements such as O, Mg, Al, Si, Ca, and Fe should be more abundant than minor elements like Na and K \citep{flynn_stern1996, sarantos2012}. However, for Si, Ca, Fe, and Ti, the derived upper limits are more than an order of magnitude lower than these predictions \citep{flynn_stern1996}. Additionally, \cite{berezhnoy2014} found that impact-produced Ca and Al atoms in the lunar exosphere were depleted relative to Na during the Perseid 2009 meteor shower, deviating from stoichiometric model expectations. The approximate binding energy or the cohesive energy values for major elements found in lunar minerals range from 1 to 10\,eV \citep{kittel2021} or even higher since the binding energy is dependent on the mineral bonds considered. Since EUV radiation of solar origin reaching the lunar surface typically has energies exceeding 10\,eV, this radiation can overcome the binding energies of these major elements. Consequently, EUV radiation may release these major elements from the mineral structures on the lunar surface through PSD.

The dynamics of the Na release processes are depicted in Figure \ref{fig:illustration}, which highlights the non-thermal mechanisms responsible for Na ejection from the lunar surface. Some Na atoms, released through momentum transfer via MIV and particle sputtering, follow ballistic trajectories under the influence of gravity. These atoms can either re-adsorb onto the lunar surface or, if they reach escape velocity, leave the Moon entirely, forming the sodium tail. We proposed that EUV and FUV radiation contribute to the release of Na from surface-bound minerals, as other non-thermal processes are known to create adsorbed Na layers. Beyond the release of Na from mineral-bound states, atoms in the adsorbate layer can be re-released through PSD, triggered by EUV and FUV radiation from the Sun. This process is shown in Figure \ref{fig:illustration}. The correlations presented in Figures \ref{fig:EUV_photonflux_Optical} and \ref{fig:GOES-VBT_comparison} provide observational evidence supporting this process. While MIV and sputtering primarily generate higher-altitude, gravitationally escaping Na populations, the higher energy from EUV and FUV radiation may also contribute to the production of such populations through PSD, potentially leading to Na atoms escaping lunar gravity.

In addition to EUV and FUV radiation, NUV may also contribute to Na release. While previous studies indicate that visible and near-ultraviolet photons ($\lambda\,>$3000\,\text{\AA}, photon energy $<$4\,eV) cause little to no detectable desorption of Na \citep{Wurz2022}, we observe a correlation between the total optical line flux from Na and the NUV irradiance. This suggests that NUV flux can induce the desorption of Na from the lunar surface, explaining the observed correlation. For UV photons with energies greater than 4\,eV, the cross-section for PSD increases substantially \citep{Yakshinskiy1999}, reaching approximately $(3\pm1)\times10^{-20}$\,$cm^{2}$ for photon energies around 5 eV. Furthermore, \cite{Mandey1998} suggests that photon energies exceeding 15–20 eV can also support the PSD process, which our observations confirm. The impulsive phase of solar flares, marked by strong UV and EUV continuum emission enhancements, could significantly influence PSD \citep{Wurz2022, Dominique2018}. While \cite{Yakshinskiy1999, Wurz2022} highlight the importance of solar UV in the 1000 to 3000\,\text{\AA} range for PSD, our analysis shows that UV photons with wavelengths shorter than 1000\,\text{\AA} also contribute to the PSD of Na on the lunar surface. It was observed that the photon flux decreases with increasing UV energy, which may reduce the amount of Na released at higher UV energies compared to lower UV energies.

The current PSD model given by \cite{Wurz2022} and \cite{Yakshinskiy1999} suggests that the PSD flux is expected to increase linearly with an increase in incident photon flux. \cite{Wurz2022} provide Equation \ref{eqn:psd_model} as a relation to estimating the flux of released Na atoms ($\Phi_{PSD}$) from the surface by PSD as a function of $\Phi_{EUV}$, the EUV flux reaching the surface.

\begin{equation}\label{eqn:psd_model}
    \Phi_{PSD} \approx \frac{1}{4}fN_{s}\Phi_{EUV}Q
\end{equation}

In Equation \ref{eqn:psd_model}, f is the fraction of the species of Na in the regolith, $N_{s}$ is the surface density of the lunar regolith, which is estimated to be a constant, $7.5 \times 10^{14}$ $cm^{-2}$ \citep{Wurz2022}. The PSD cross-section (Q) is estimated to be a constant for the 250 - 400\,nm wavelength range and is estimated by \cite{Yakshinskiy1999} as $(3\pm1)\times10^{-20}$\,$cm^{2}$, and they also indicate that the cross-section increases rapidly beyond UV photon energy of $\sim$5\,eV, but is expected to be a constant for a particular wavelength. Hence, $\Phi_{PSD}$ is expected to vary linearly with $\Phi_{EUV}$ for a given wavelength. In contrast, our data exhibits a non-linear variation. The measured Na brightness in the lunar exosphere increases non-linearly with photon flux across the FUV to EUV wavelength range. This variation could be influenced by factors such as observation altitude, position angle, phase of the Moon and changes in the g-value with different $T_{ch}$ in our data. However, the effect of altitude remains within the uncertainties of our Na I brightness measurements. The impact of position angle variation is minimal, as observations were consistently taken vertically above the same region each night. Additionally, variations in the g-value along the lunar orbit were evident in our data. These variations, along with changes in the measured $T_{ch}$, were incorporated into the estimation of $N_{los}$. The $N_{zen}$ was then derived from $N_{los}$ using the Chamberlain model. To reduce the influence of altitude and position angle variations within a single night of observation, daily-averaged values were used in the analysis. Importantly, the observed non-linear correlation persisted even after applying these corrections and averaging procedures, indicating the robustness of the observed trend between the UV photon flux and the Na\,I brightness measurements. The optical Na brightness measurements also suggest a non-thermal population based on the $T_{ch}$. The deviation between sodium abundance and a linear dependence on EUV flux indicates that the flux may also include contributions from other processes, such as thermal desorption, sputtering or MIV, in addition to PSD. Among these, MIV is expected to peak on the leading side of the Moon in its orbital movement around the Earth \citep{pokorny2019}. During the first quarter (waxing phase), the Moon's velocity vector positions the leading side at the western equatorial region, which is in shadow. We observe only the brighter eastern equatorial region at the apparent sub-solar point. Additionally, the sporadic meteor shower periods do not significantly align with our observations \citep{kronk2013meteor}. Therefore, we assume that the contribution of micro-meteorite impact vaporization to the observed Na flux is minimal. The non-linear trends we see are likely caused by sputtering, the other non-thermal process.

The strong presence of the non-thermal components was confirmed by the $T_{ch}$ estimated in our observations, which exceeds 3000\,K. \cite{killen2019} found temperatures ranging from $\sim$2255\,K to 6765\,K at altitudes extending up to 1 lunar radius. \cite{killen2021} consistently measured temperatures around 4500\,K on average, ranging from 2250\,K to 9000\,K. \cite{mendillo1993} observed equatorial temperatures of $\sim$4500\,K from waxing limb data. \cite{Kuruppuaratchi2018} also observed linewidth-derived temperatures ranging from 2500\,K to 9000\,K. They also noted high source concentrations near the sub-solar point. Resonant scattering measurements of lunar Na by the SELENE (Kaguya) orbiter also indicate temperatures ranging from 2400 to 6000\,K \citep{kagitani2010}. \cite{killen2021} studied the exosphere between altitudes 143\,km to 1738\,km. The scale height measured at these heights ranges between 500 to 2000\,km. The values in this work also match within this range except on 16 and 17 March 2024, where we observe high $T_{ch}$. Our observations within 590\,km above the surface at the apparent sub-solar point also indicate similar temperatures, indicating a contribution from the non-thermal processes in releasing the Na we observe in the lunar exosphere. \cite{killen2019} and \cite{killen2021} also indicate that their observations within one Lunar radius from the lunar surface likely captured emissions from a predominant hot source or the higher temperatures of a distribution function within the exosphere. Likewise, our observations may also detect this component. Therefore, a possible contribution from sodium atoms released via sputtering by solar wind protons cannot be ruled out. To investigate this, we analysed solar wind parameters—including speed, proton density, and proton temperature—sourced from NASA/GSFC’s OMNI 1-minute dataset via OMNIWeb \citep{King_Papitashvili_2020} for the period of our observations. The solar wind speed varied between 336.3 and 397.3 km\,s$^{-1}$, the proton temperature ranged from 27,025 K to 105,759 K, and the proton density spanned 2.37 to 8.31 cm$^{-3}$. Correlation analysis between these parameters and the measured $N_{zen}$ showed variations within the measurement uncertainties. Thus, the contribution from sputtering due to the solar wind particles appears minimal and unlikely to influence the observed deviation from linearity in this PSD-dominated system.


\section{Conclusions}\label{sec:conclusion}
We carried out high-resolution optical spectroscopy of the lunar exosphere during the first quarter from January to March 2024 using VBT. The simultaneous EUV and FUV flux of solar origin was obtained from the spectral line fluxes of GOES-r along with a daily averaged NUV flux from TSIS-1. Various correlations between these datasets were studied, and arrive at the following conclusions:
 
\begin{enumerate}
\item The optical line flux from the Na D lines, which reflects the sodium abundance in the lunar exosphere, correlates with the EUV and FUV fluxes reaching the Moon from the Sun. Our study reveals a non-linear increase in Na abundance with increasing EUV and FUV flux within the wavelength range of 256 to 1405,\text{\AA}. This relationship is observed for the first time using simultaneous measurements from VBT and GOES. The Pearson correlation test yielded a p-value of $3.31\times10^{-4}$ with a correlation coefficient of 0.834, while the Spearman rank correlation test resulted in a p-value of $5.28\times10^{-6}$ and a correlation coefficient of 0.956. These values indicate a strong and statistically significant positive correlation. The potential effects of observation altitude are found to be within the measurement uncertainties, while the impact of position angle is minimal. Our findings confirm that EUV radiation with energy $>$8.8\,eV, originating from the Sun, contributes to the release of Na from the lunar surface into the exosphere via PSD.

\item The mean $N_{los}$ and $N_{zen}$ derived from our observations are $5.6 \times 10^{9}$\,$atoms\,cm^{-2}$ and $3.3 \times 10^{9}$\, $atoms\,cm^{-2}$, respectively. Variations in the $g$-value, driven by changes in $RV_{hc}$ and exospheric temperature, were incorporated into the column density estimates. The persistence of the non-linear correlation in daily-averaged values suggests that this trend is robust against potential influences from altitude and position angle variations. These results indicate that the non-linearity with EUV flux remains significant despite fluctuations in both $RV_{hc}$ and exospheric temperature along the lunar orbit.

\item The measured characteristic temperature ($T_{ch} \sim 6700\,\mathrm{K}$) and corresponding scale height ($\sim1500\,\mathrm{km}$) indicate the presence of non-thermal components in the lunar Na exosphere, consistent with non-thermal release mechanisms. These estimates are in agreement with previous reports. Growth factors derived from the observed non-linearity exhibit a decreasing trend with increasing photon energy from FUV to EUV, mirroring the decline in solar photon flux across this spectral range during all observation nights. This reduction in photon flux likely contributes to the decrease in the growth factor. The pronounced deviation from linearity potentially indicates enhanced PSD efficiency at EUV wavelengths within 256--304\,\AA. However, a contribution from other thermal or non-thermal processes is expected.

\item The observations in this work indicate that UV photons in the EUV range can effectively release Na from the lunar surface, contributing to the Na exosphere. The observed EUVS wavelengths have photon energies above 8.8\,eV and could possibly break mineral bonds to release bounded Na from minerals, in addition to the release of Na found as an adsorbate layer.

\item PSD at EUV wavelengths could potentially contribute to the gravitationally escaping Na population and the higher-altitude Na in the exosphere. In addition, EUV radiation may overcome the binding energies of various elements, releasing bounded major elements like O, Mg, Al, Si, Ca, and Fe, which are present in higher abundance on the Lunar surface than minor elements like Na and K.

\item The lower limit of $N_{zen}$ of Fe\,I in the lunar exosphere was estimated to be $3.9\times10^{10}$ $atoms\,cm^{-2}$. The lower limit is consistent with the estimates from \cite{berezhnoy2014} and \cite{flynn_stern1996}.

\item The solar NUV flux positively correlates with the observed optical line flux from the Na D lines in the exosphere. This provides observational evidence for the involvement of NUV in releasing Na from the lunar surface, where previous literature suggested that visible and NUV photons (photon energy $<$4\,eV) cause little or no detectable desorption of Na.

\item An increased $T_{ch}$ was observed during the measurements on 16 and 17 March 2024, coinciding with multiple solar flaring events near the observation times (UTC). This elevated temperature may be attributed to enhanced solar activity. The observed non-linear correlation may thus be influenced by additional processes such as thermal desorption, sputtering, or MIV. However, the solar wind parameters during the observations--speed (336.3--397.3\,km\,s$^{-1}$), proton temperature (27,025--105,759\,K), and proton density (2.37--8.31\,cm$^{-3}$)--exhibited no significant correlation with the measured $N_{zen}$ within the limits of observational uncertainty.

\end{enumerate}
\noindent
The role of solar EUV in PSD from the lunar surface is evident from our observations. The non-linear correlation, deviating from the existing understanding of a linear relationship, suggests the possibility of a modified PSD model for the EUV regime. Additionally, laboratory experiments are needed to better constrain the PSD cross-section in the EUV range.

\section*{Acknowledgements}
We would like to acknowledge the referee for the insightful comments and constructive feedback, which significantly improved the quality and clarity of this article. We extend our sincere gratitude to the staff of the Technical Departments at Vainu Bappu Observatory (VBO) for their invaluable support in modifying the Vainu Bappu Telescope (VBT) setup, essential for the observations in this study. Special thanks go to the observing team - Surendranath, Vimal Kumar, Umayal, S. Ramesh, and Bhaskar for their exceptional assistance. We are also grateful for insightful discussions with Dr. Sankarasubramanian K (ISRO) and Dr. Savithri H. Ezhikode, which greatly helped in designing the observations and analysis. We thank the Indian Institute of Astrophysics for providing necessary observational facilities and the Indian Space Research Organization (ISRO) for funding this research under the Chandrayaan-2 AO project (sanction no: DS\_2B-13013(2)/8/2022-sec.2). SSK and AKR acknowledge the financial support from CHRIST (Deemed to be University, Bangalore) through the SEED money project (No: SMSS-2220, 12/2022). We acknowledge the facility support from the FIST program of DST (SR/FST/PS-I/2022/208). Additionally, we appreciate the research support provided by CHRIST (Deemed to be University) throughout this work.

\section*{Data Availability}
The data underlying this article will be shared on reasonable request to the corresponding author.

\bibliographystyle{mnras}
\bibliography{references} 

\bsp	
\label{lastpage}
\end{document}